\documentclass[journal]{IEEEtran}
\usepackage[utf8]{inputenc}
\usepackage{comment}
\usepackage{amsmath,amsmath,bm,amsfonts}
\usepackage{graphicx}
\usepackage{textcomp}
\usepackage{booktabs}
\usepackage[caption = false]{subfig}
\usepackage{cite}
\usepackage{todonotes}
\usepackage{multirow}

\newcommand{\zvi}[1]{\mathbf{z}_{#1}}
\newcommand{\svi}[1]{\mathbf{s}_{#1}}
\newcommand{\cvi}[1]{\mathbf{c}_{#1}}

\title{Unsupervised Speech Segmentation and Variable Rate Representation Learning using \\ Segmental Contrastive Predictive Coding}
\author{Saurabhchand Bhati$^{\dagger}$, Jes\'us Villalba$^{\dagger,\ddagger}$, Piotr \.Zelasko$^{\dagger,\ddagger}$, Laureano Moro-Velazquez$^{\dagger}$, Najim Dehak$^{\dagger,\ddagger}$ \\
$^{\dagger}$Center for Language and Speech Processing, Johns Hopkins University, USA \\
 $^{\ddagger}$Human Language Technology Center of Excellence, Johns Hopkins University, USA  \\
 \{sbhati1\thanks{This project was supported by NSF Award 1909075},jvillalba,pzelasko,laureano,ndehak3\}@jhu.edu}

\begin{document}

\maketitle

\begin{abstract}
Typically, unsupervised segmentation of speech into the phone and word-like units are treated as separate tasks and are often done via different methods which do not fully leverage the inter-dependence of the two tasks. Here, we unify them and propose a technique that can jointly perform both, showing that these two tasks indeed benefit from each other. Recent attempts employ self-supervised learning, such as contrastive predictive coding (CPC), where the next frame is predicted given past context. However, CPC only looks at the audio signal's frame-level structure. We overcome this limitation with a segmental contrastive predictive coding (SCPC) framework to model the signal structure at a higher level, e.g., phone level. A convolutional neural network learns frame-level representation from the raw waveform via noise-contrastive estimation (NCE). A differentiable boundary detector finds variable-length segments, which are then used to optimize a segment encoder via NCE to learn segment representations. The differentiable boundary detector allows us to train frame-level and segment-level encoders jointly. 
Experiments show that our single model outperforms existing phone and word segmentation methods on TIMIT and Buckeye datasets. We discover that phone class impacts the boundary detection performance, and the boundaries between successive vowels or semivowels are the most difficult to identify. Finally, we use SCPC to extract speech features at the segment level rather than at uniformly spaced frame level (e.g., 10 ms) and produce variable rate representations that change according to the contents of the utterance. We can lower the feature extraction rate from the typical 100 Hz to as low as 14.5 Hz on average while still outperforming the MFCC features on the linear phone classification task.
\end{abstract}


\section{Introduction}
Voice-enabled interfaces for human-machine interaction have made significant progress in recent years~\cite{baevski2020wav2vec,oord2016wavenet,synnaeve2019end,wang2019transformer}. Most of the success can be attributed to very deep neural networks trained on abundant resources, including thousands of hours of transcribed data. However, for most spoken languages worldwide, e.g., regional languages, vast amounts of labeled data are not available. Zero Resource speech processing aims to develop alternate techniques that can learn directly from data without any or minimal manual transcriptions~\cite{versteegh2015zero,dunbar2017zero,dunbar2019zero}. It has several applications, including preserving endangered languages, building speech interfaces in low-resource languages, and developing predictive models for understanding language evolution~\cite{rasanen2012computational}.

Unsupervised discovery of phone or word-like units forms the core objective of Zero Resource speech processing~\cite{jansen2011efficient,badino2014auto,huijbregts2011unsupervised,lee2012nonparametric,siu2014unsupervised,kamper2017segmental,bhati2017unsupervised,kamper2017embedded,bhati2018phoneme}. The representation techniques employed to characterize the speech signal dictate the quality of the segmentation and clustering and, in turn, the entire process of discovering linguistic units from the speech signal. 
The speech signal contains information about multiple factors, such as linguistic units~\cite{hinton2012deep}, speakers~\cite{dehak2010front}, emotion~\cite{kwon2003emotion}, and others. 
In the unsupervised scenario, we lack the guidance of manual transcriptions to select the relevant features and marginalize irrelevant information. 
A good speech representation becomes crucial for unsupervised systems' good performance. 
Self-supervised methods have emerged as a promising technique for representation learning from unlabeled speech data~\cite{oord2018representation,schneider2019wav2vec,baevski2020wav2vec,kreuk2020self,devlin2018bert,liu2019roberta,chen2020simple}.  
In the self-supervised learning (SSL) scenario, the `self' part refers to the generation of pseudo-labeled training data for an auxiliary task, and `supervised' refers to the supervised training of the underlying neural model. 
SSL has been shown to be effective for learning representations in natural language processing~\cite{devlin2018bert,liu2019roberta}, images~\cite{chen2020simple}, and speech processing~\cite{oord2018representation,schneider2019wav2vec}. 
While most SSL work in speech focuses on learning representations~\cite{oord2018representation,schneider2019wav2vec,baevski2020wav2vec}, SSL has recently been used to identify the spectral changes from the raw waveform and detect phone boundaries~\cite{kreuk2020self}. 

The SSL framework exploits the temporal structure present in the speech to learn latent representations. For instance, in~\cite{kreuk2020self}, the auxiliary task is to identify the next frame's latent representation given the latent representation of a reference frame. A feature extractor, e.g., Convolutional Neural Network (CNN), that maps the speech signal to a latent space is optimized using the Noise Contrastive Estimation (NCE)~\cite{gutmann2010noise} to correctly identify the next frame within a set of frames that includes random distractor frames. 
In contrast, the speech signal is composed of underlying linguistic units like phones or words instead of frames.
Thus, it would be beneficial to learn from the speech structure at the phone level and use that information to predict the next frame or even segment. 
But unsupervised phone segmentation is not straightforward due to variable phone length, which is why current SSL frameworks work at the frame level. 

The semantic content in the audio is not uniformly spaced. It varies over time depending upon the content and context. Feature extraction methods like MFCC process the audio in uniformly spaced chunks and compute acoustic feature vectors at equally spaced time intervals (10 ms, typically). This leads to redundancy in the successive feature vectors, as adjacent vectors often have the same phonetic class. The same thing happens while extracting features for silence, where adjacent frames add little or no extra information. One second of speech is represented by 100 frames regardless of the number of phones or silence regions. In contrast to fixed-rate methodologies, extracting one feature per phone level could reduce the number of features extracted per second on average. Phone boundaries mark the changes in the underlying phones and thus the changes in underlying semantic content. Therefore, it would be beneficial to incorporate unsupervised phone boundary detection as part of the learning process. We can sample the speech whenever there is a segment change and extract a single representation per segment. An ideal segment-level representation would match the performance of frame-level representation on downstream tasks, e.g., the phone classification task.

In this work, we extend our Segmental Contrastive Predictive Coding (SCPC)~\cite{bhati2021segmental} method for unsupervised phone and word segmentation. SCPC can exploit the structure in speech signals at a high scale, i.e., phones, during the learning process. Here, we analyze the impact of various model hyper-parameters and study where SCPC works better or worse than prior works. 
The boundary detector relies on a manually tuned threshold for segmentation. Here we also learn the boundary threshold along with model parameters instead of manually fixing it. In the end, we present experiments on variable rate representation learning with SCPC and use SCPC to extract representations at the segmental level, e.g., one representation per segment (phone).

In this approach, we start from the raw waveform and train an encoder via the next frame prediction to extract frame-level latent representations. We then use a differentiable boundary detector to extract variable-length segments. We develop a simple yet efficient differentiable boundary detection that can detect boundaries without any constraints on segment length or the number of segments in an utterance. Then, we obtain the segment level representation by encoding the averages of the variable-length segments through a segment encoder. The differentiable boundary detector allows information to flow between frame and segment encoder. The model is trained in an end-to-end manner.

A word can be thought of as a sequence of segments. In natural languages, segments in a word tend to have higher prediction probability, i.e., it is easier to predict the next segment given the past segment than the segments across word boundaries~\cite{harris1954distributional,saffran1996word}. Statistical word segmentation methods often rely on the prediction probabilities between subword units, e.g., syllables or segments~\cite{goldwater2009bayesian}. Infants seem to be able to exploit this tendency of natural languages to divide continuous speech into word-like units~\cite{aslin1998computation}. Here, we rely on the prediction probability between segments as a cue for word segmentation. By locating the time-points with low prediction probability between adjacent segments are considered candidates for word boundaries.

Our proposed methods enable boundary detection and segment representation computation in a batch manner for faster training. At the frame level, the model is optimized to predict the next frame. At the segment level, the auxiliary task becomes the next segment prediction.
Unlike frames, for segments, the previous segment might not be enough to predict the next segment. We need to learn the context or the order in which the segments occur. We use a recurrent neural network to capture the segment context.
The joint training allows the model to capture the structure present in the speech at multiple levels, i.e., frames and phones level, and these two mutually benefit from each other.
We evaluate our proposed methods on TIMIT~\cite{garofolo1993timit} and Buckeye~\cite{pitt2005buckeye} datasets for phone and word boundary detection. Our proposed method outperforms state-of-the-art phone and word segmentation methods. %
In this study, our experiments show that SCPC based segment-level representations outperform the frame-level features, i.e., MFCC on linear phone classification task on TIMIT database, while significantly reducing sampling rate.

The rest of the paper is organized as follows: Section 2 refers to the relevant prior work, Section 3 describes the SCPC model in detail, Section 4 provides details about dataset, evaluation metric and provides experimental results, Section 5 analyzes the segments learned by CPC and SCPC based segmentation methods. Section 6 describes a SCPC based variable-rate representation learning system. Section 7 concludes the paper with a discussion and future work.

\section{Related Work }
Most previous works reduce the phone boundary detection task to a boundary classification task at each time step.
Michel et al.~\cite{michel2016blind} use HMM or RNN to predict the subsequent frame. A peak detection algorithm identifies the high prediction error regions, which are then used as phone boundaries. Wang et al.~\cite{wang2017gate} train an RNN autoencoder and track the norm of intermediate gate values over time, e.g., forget gate for LSTM. A peak detection algorithm identified the boundaries from the gate norm over time.  Kreuk et al.~\cite{kreuk2020self} train a CNN encoder to distinguish between adjacent frame pairs and random pairs of distractor frames. To detect the phone boundaries, a peak detection algorithm is applied over the model outputs. This method achieves state-of-the-art phone segmentation on TIMIT and Buckeye dataset. All these methods try to exploit the structure at frame level to detect phone boundaries, whereas SCPC also looks at higher-level structure, i.e., segment level, along with the frame-level structure to detect phone boundaries.    

Word segmentation is an important problem in Zero resource speech processing. 
Some studies employ Bayesian Segmental GMM~\cite{kamper2017segmental} and Embedded Segmental K-Means (ES-KMeans)~\cite{kamper2017embedded} that start from an initial set of subword boundaries and then iteratively eliminate some of the boundaries to arrive at frequently occurring longer word patterns. The initial subword boundaries are not adjusted during the process. As a result, the performance of the ES-Kmeans critically depends on the initial boundaries~\cite{bhati2018phoneme}. 
Unsupervised word segmentation techniques that can discover word-like units from space-removed sequences of characters have been used to discover words from sequences of discovered acoustic units from speech~\cite{bhati2017unsupervised}. Kamper et al.~\cite{kamper2020towards} proposed methods to constrain the outputs of vector-quantized~(VQ) neural networks and assign blocks of consecutive feature vectors to the same segment. 
First, vector quantized variational autoencoder~(VQ-VAE)~\cite{chorowski2019unsupervised} and vector quantized contrastive predictive coding~(VQ-CPC)~\cite{baevski2019vq} models are trained to encode the speech signal into discrete latent space. Then, dynamic programming~(DP) is used to merge the frames to optimize a squared error with a length penalty term to encourage longer but fewer segments. 
The DP segmentation output is further segmented using word segmentation algorithms, e.g., adopter grammar~\cite{johnson2007adaptor}, or the Dirichlet process model~\cite{goldwater2009bayesian},  typically used for segmenting sequences of space-removed characters or phones. These algorithms have multiple stages without feedback between the stages, e.g., the DP segmentation remains fixed while applying the word segmentation step. SCPC reduces the number of steps required for word segmentation and allows feedback between different components. 

The tasks of phone segmentation and word segmentation are often done via separate methods~\cite{kamper2017embedded,kamper2017segmental,bhati2018phoneme,bhati2019unsupervised}. Some methods use phone segmentation as a starting point for word boundaries~\cite{bhati2017unsupervised,bhati2018phoneme,bhati2019unsupervised}. However, none of the methods jointly do phone and word segmentation. Here, we propose a single system capable of doing both and able to exploit the two tasks' inter-dependencies. 

On the other hand, self-supervised techniques have greatly improved the quality of speech representation. ASR system built on top of these representations requires much less data to achieve similar error rates~\cite{baevski2020wav2vec}. However, the CPC representations are extracted at a fixed sampling rate. Supervised ASR models can handle the uniformly sampled acoustic features and employ variable rate data processing to match the phone or word output which is sampled at a much slower rate. e.g., HMMs classify several acoustic frames into a single phone. Neural transducers e.g. Connectionist Temporal Classification (CTC)~\cite{graves2006connectionist}, RNN-Transducer~\cite{graves2012sequence} can collapse the adjacent features to align inputs and shorter outputs. 
However, these models require ground-truth transcriptions for training.

There have been several attempts to impose slow changes~\cite{wiskott2002slow} on the representations learned in an unsupervised manner. In~\cite{chorowski2019unsupervised} a time-jitter regularization was proposed to reduce the variability between adjacent embeddings of VQ-VAE. Authors of~\cite{chorowski2019unsupervised2} added a penalty to divide the VQ-VAE latent codes in a given number of piecewise-constant pieces. The study~\cite{kamper2020towards} generalized this approach and proposed dynamic programming based approach to obtain phone segmentation from VQ-VAE and VQ-CPC features. SCPC inherently generates piece-wise constant representation, i.e., the representation change abruptly at segment boundaries and is relatively similar within segments. 

\section{Segmental Contrastive Predictive Coding}
\begin{figure}
    \centering
    \hspace*{-0.04\linewidth}
    \includegraphics[width=3.3in]{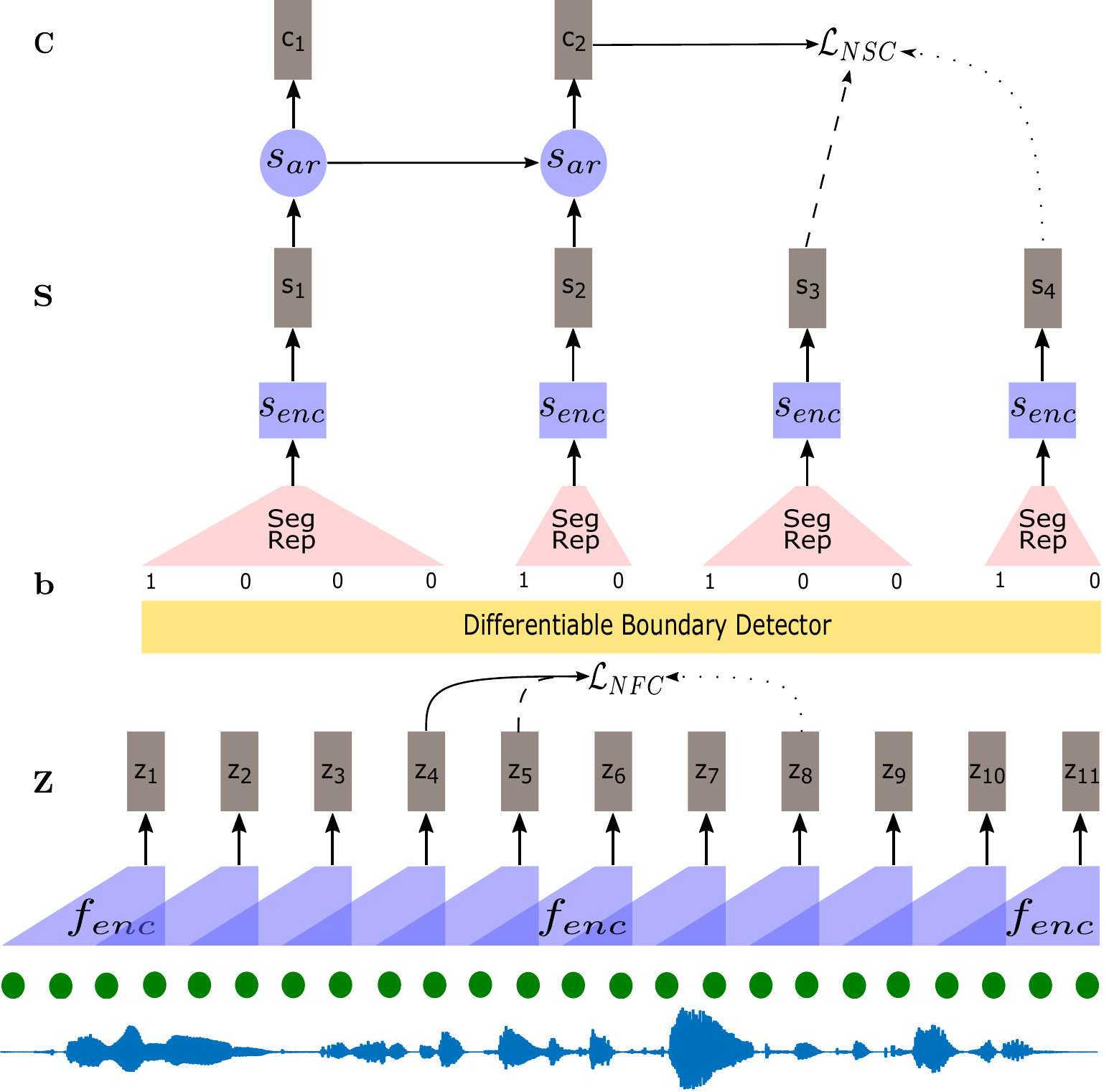}
    \vspace{-4mm}
    \caption{Overview of the Segmental Contrastive Predictive coding architecture. The solid line represents the reference frame (segment) $\zvi{4} (\cvi{2})$, the dashed line represents the positive  frame (segment) $\zvi{5} (\svi{3})$ and the dotted line represent the negative distractor randomly sampled from the signal}
    \label{fig:SCPC}
    \vspace{-6mm}
\end{figure}

We train the SCPC system, depicted in Figure~\ref{fig:SCPC}, by solving contrastive tasks at multiple scales: at the frame level, with next frame classification (NFC) loss $ \mathcal{L}_{\mathrm{NFC}}$, and at the segment level, with next segment classification (NSC) loss $\mathcal{L}_{\mathrm{NSC}}$, which require the model to identify the true latent representation within a set of distractors. The system's final objective is to minimize the overall loss ($ \mathcal{L}$), composed of both the next frame prediction and the next segment prediction losses:
\begin{equation}
    \mathcal{L} = \mathcal{L}_{\mathrm{NFC}} + \mathcal{L}_{\mathrm{NSC}} \;
\end{equation}

\subsection{Next Frame Classifier}

Let the sequence $\mathbf{X} = (x_1,x_2,...,x_T)$ represent a speech waveform.
We learn an encoding function $f_{\mathrm{enc}}:\mathbf{X}\rightarrow\mathbf{Z}$ (more details in section ~\ref{Model_arch})that maps audio sample sequences to latent spectral representations, $\mathbf{Z}(\in \mathbb{R}^{ p \times L}) = (\zvi{1},\zvi{2},...,\zvi{L})$ at lower frequency. Each p-dimensional vector $\zvi{i}$ corresponds to a 30 ms audio frame extracted with 10 ms shift. Given frame $\zvi{t}$, the model is trained to identify the next frame $\zvi{t+1}$ correctly within a set of $K+1$ candidate representations $\Tilde{\zvi{}} \in \mathcal{Z}_{t}$, which includes $\zvi{t+1}$ and $K$ distractors--randomly sampled from the same utterance, as
\begin{equation}
    \mathcal{L}_{\mathrm{NFC}} = -\log \frac{\exp(\mathrm{sim}(\zvi{t},\zvi{t+1}))}{\sum_{\Tilde{\zvi{}} \in \mathcal{Z}_{t} } \exp(\mathrm{sim}(\zvi{t},\Tilde{\zvi{}} ))}
\end{equation}
where $\mathrm{sim}(\mathbf{x},\mathbf{y}) =  \frac{\mathbf{x}\mathbf{y}^{T}}{\Vert \mathbf{x} \Vert \Vert \mathbf{y} \Vert} $ denotes the cosine similarity.

\subsection{Differentiable Boundary detection}
A phone segment can be thought of as a sequence of frames. 
The frames that follow each other with high similarity are likely to be from the same segment. 
If there is a high dissimilarity between adjacent frames, then it might indicate a segment change. $\mathbf{d}=(d_1,d_2,...,d_{L-1})$ captures the dissimilarity between two adjacent frames,
\begin{equation}
\begin{split}
& \mathbf{d} = 1 - (\mathbf{d^{s}} - \mathrm{min}(\mathbf{d^{s}})) / (\mathrm{max}(\mathbf{d^{s}}) -  \mathrm{min}(\mathbf{d^{s}})) \\
 \end{split}
\end{equation}
where $\mathbf{d^s} = (d^s_1,d^s_2,...,d^s_{L-1})$ s.t. $ d^{s}_{t} = \mathrm{sim}(\zvi{t},\zvi{t+1}) $  captures similarity between adjacent frames.
To locate the segment boundaries, we need to find the frame indexes with high dissimilarity values. We can do that by finding peak locations in $\mathbf{d}$. 
Equation~\ref{eq:peaks} defines the peak detectors $\mathbf{p^{(1)}}$, $\mathbf{p^{(2)}}$ and $\mathbf{p}$ whose $t^{th}$ entries are
\begin{equation}
\label{eq:peaks}
\begin{split}
&    p_t^{(1)} = \min(\max(d_{t} -  d_{t+1},0), \max(d_{t} -  d_{t-1},0)) \\
&    p_t^{(2)} = \min(\max(d_{t} -  d_{t+2},0), \max(d_{t} -  d_{t-2},0)) \\
&    p_t = \min(\max(\max(p_t^{(1)},p_t^{(2)}) - \mathrm{thres},0), p_t^{(1)})
\end{split}    
\end{equation}
$p_t^{(1)}$ captures if $d_t$ is greater than both $d_{t-1}$ and $d_{t+1}$ or not. It will be zero if it is smaller than either $d_{t-1}$ or $d_{t+1}$ and non-zero otherwise. By itself, $p_t^{(1)}$ might be noisy and over-segment. We amend that by introducing $p_t^{(2)}$, which compares $d_t$ with $d_{t-2}$ and $d_{t+2}$. A peak in the low range can be more informative than one that is higher but otherwise is a trivial member of a tall range. So we check if the height of the peak as compared to its neighbors is more than a threshold (boundary threshold) or not instead of the peak height directly.

The vector $\mathbf{p}$ will have zeros wherever there are no boundaries and non-zero if there is a boundary. The non-zeros are values different across peaks, so we scale them to make all the non-zero values consistently $1$. This is also helpful in vectorizing the segmentation representation process (see section C). We can do that by taking $\tanh$ of a large scalar multiple e.g. 100 of $p$, but this might result in vanishing gradients. We use a gradient straight-through estimator~\cite{oord2017neural,bengio2013estimating} %
for obtaining the boundary variables, 
\begin{equation}
\label{eq:bounds}
\begin{split}
&    \mathbf{b_\mathrm{soft}} = \tanh(10\;\mathbf{p}) \\
&    \mathbf{b_\mathrm{hard}} = \tanh(1000\;\mathbf{p}) \\
&    \mathbf{b} = \mathbf{b_\mathrm{soft}} + 
\mathrm{sg}(\mathbf{b_\mathrm{hard}}-\mathbf{b_\mathrm{soft}})
\end{split}
\end{equation}
where $\mathrm{sg}$ is the stopping gradient function that avoids gradients to flow through $\mathbf{b}_\mathrm{hard}$, which has exploding gradient at 0.
The $\mathbf{b}$ is $1$ at boundaries and $0$ elsewhere. This method can detect segments in a batch manner without any assumptions on segment length or the number of segments.

\begin{figure}
    \centering
    \includegraphics[width=1.05\columnwidth]{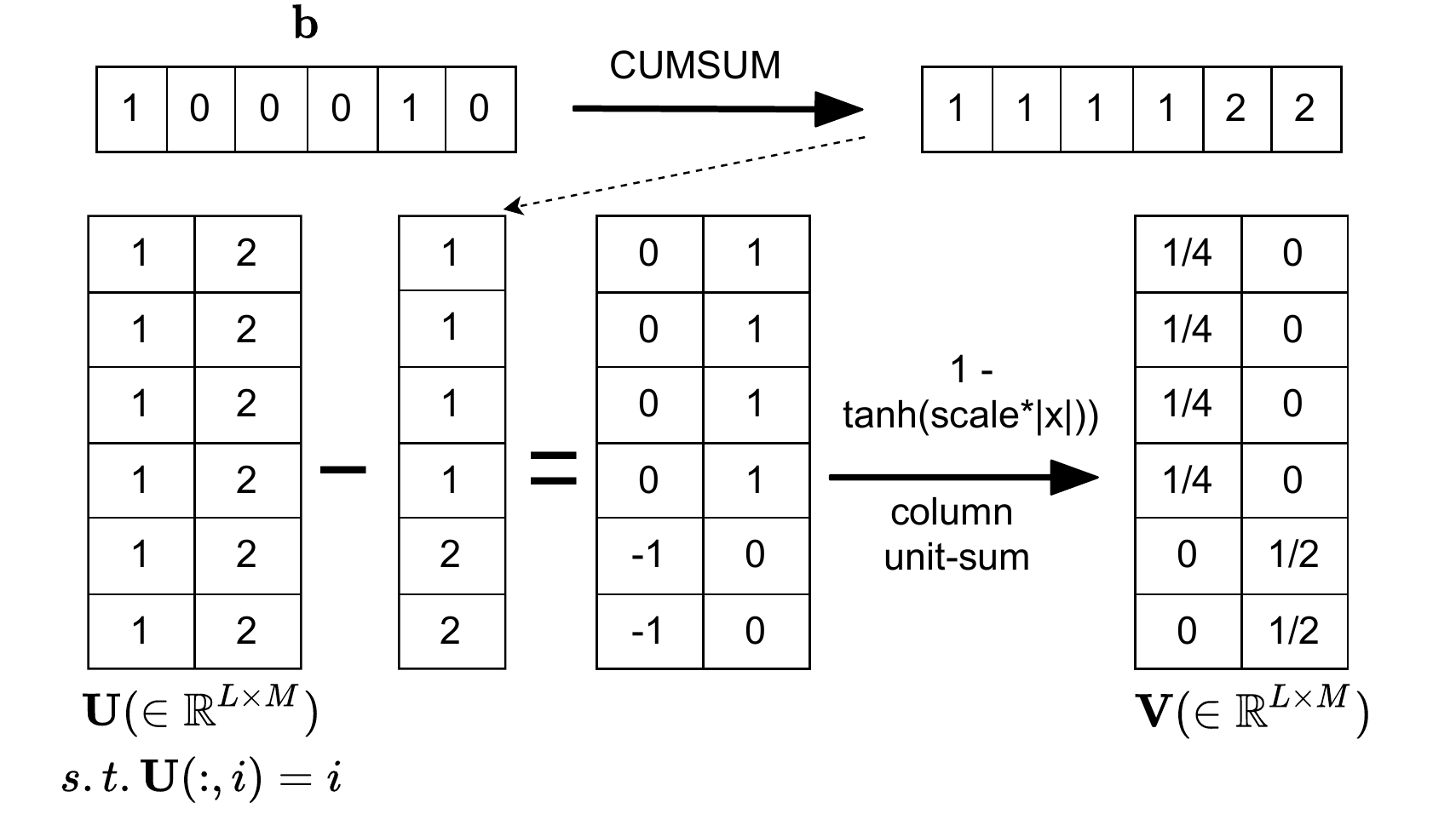}
    \caption{Weight generation for obtaining segment means in a differentiable manner. The number of segments $M$ can be computed by taking the sum of $\mathbf{b}$.}
    \label{fig:mu_gen}
    \vspace{-5mm}
\end{figure}

\subsection{Segment representations}
After the boundary detection step, the feature sequence $\mathbf{Z} = (\zvi{1},\zvi{2},...,\zvi{L})$ is segmented into disjoint contiguous segments $\mathbf{S} = (\svi{1},\svi{2},...,\svi{M})$. 
The next step is to obtain representations for these variable-length segments. A fixed-size representation for segments would be easier to work with and would allow different segments to be compared straightforwardly. We experiment with the following representations for a variable-length segment.
\subsubsection{Segment average}
Since we decide a segment boundary based on high dissimilarity between frames, the frames that lie within a segment will be close to each other. We represent a segment via the average of the constituting frames and feed the averages through a segment encoder $\mathrm{s_{\mathrm{enc}}}$ to generate the segment representations. From Figure~\ref{fig:SCPC}, we see that the first four frames, $\zvi{1:4}$ belong to the first segment $\svi{1}$, and the next two frames $\zvi{5:6}$ belong to the second segment $\svi{2}$. The segment representations are given as $\svi{1} = \mathrm{s_{\mathrm{enc}}} (\frac{1}{4} \sum_{i=1}^{4} \zvi{i} )$, 
$\svi{2} = \mathrm{s_{\mathrm{enc}}} ( \frac{1}{2} \sum_{i=4}^{5} \zvi{i} ) $. 
We could iteratively compute those averages segment by segment, but that would be prohibitively slow. Instead, we propose to vectorize those computations
following the steps in Figure \ref{fig:mu_gen}. 

The vector $\mathbf{b}$ contains $1$ at segment boundaries and $0$ otherwise. We compute the cumulative sum of the $\mathrm{b}$ from left to right, which tells us which segment $i^{th}$ frame belongs to. Then we generate a weight matrix $\mathbf{W}(\in \mathbb{R}^{L\times M})$ such that $\mathbf{W}(:,i) = i$. The number of segments $M$ can be computed by taking the sum of $bounds$. From each column of the weight matrix we subtract cumulative sum of the $\mathrm{bounds}$ and pass it through the function $1 - \mathrm{tanh(scale*\vert x \vert)}$. Making the column sum to $1$ gives us the desired weight matrix. Segment representation can be obtained by multiplying the $\mathbf{Z}(\in \mathbb{R}^{p \times L})$ and $\mathbf{W}(\in \mathbb{R}^{L\times M})$ and feeding it through $\mathrm{s_{\mathrm{enc}}}$. 

\subsubsection{Segment middle point}
The frame representation of middle point of the segment is used as the segment representation. For e.g. for the first segment which contains frames 1 to 4, the segment representation is equal to the frame representation of the 2 frame i.e. $\svi{1} = \mathrm{s_{\mathrm{enc}}} (\zvi{2} )$. 

\subsubsection{Segment maximum}
The maximum of the frame representations is used as the segment representation. For e.g. for the first segment which contains frames 1 to 4, the segment representation is element wise maximum to the frame representation of the 4 frame along the time dimension i.e. $\svi{1} = \mathrm{s_{\mathrm{enc}}} (max(\zvi{1:4} ))$.

\subsubsection{Weighted Average}
In the average all the frames are weighted equally. We propose to use the self-similarity between the frames within a segment to weight the frames before taking the average.  
\begin{equation}
\begin{split}
 &    \svi{1}^{wavg} = \mathrm{s_{\mathrm{enc}}} (\mathrm{avg} (\mathbf{Z}_{1:4} * \mathrm{softmax} (\mathbf{Z}_{1:4}' * \mathbf{Z}_{1:4}) ) ) \\
 &     \svi{2}^{wavg} = \mathrm{s_{\mathrm{enc}}} (\mathrm{avg} (\mathbf{Z}_{5:6} * \mathrm{softmax} (\mathbf{Z}_{5:6}' * \mathbf{Z}_{5:6}) ) )
\end{split}
\end{equation}

After multiplying the self-similarity matrix with the features matrix, the technique mentioned above for calculating averages can be used. 

\begin{figure*}[h!]
    \centering
    \subfloat[Highest R-val, "don't do charlie's dirty dishes"]{
    \begin{minipage}{0.95\columnwidth}
    \includegraphics[ height=1.0in,width=0.95\columnwidth]{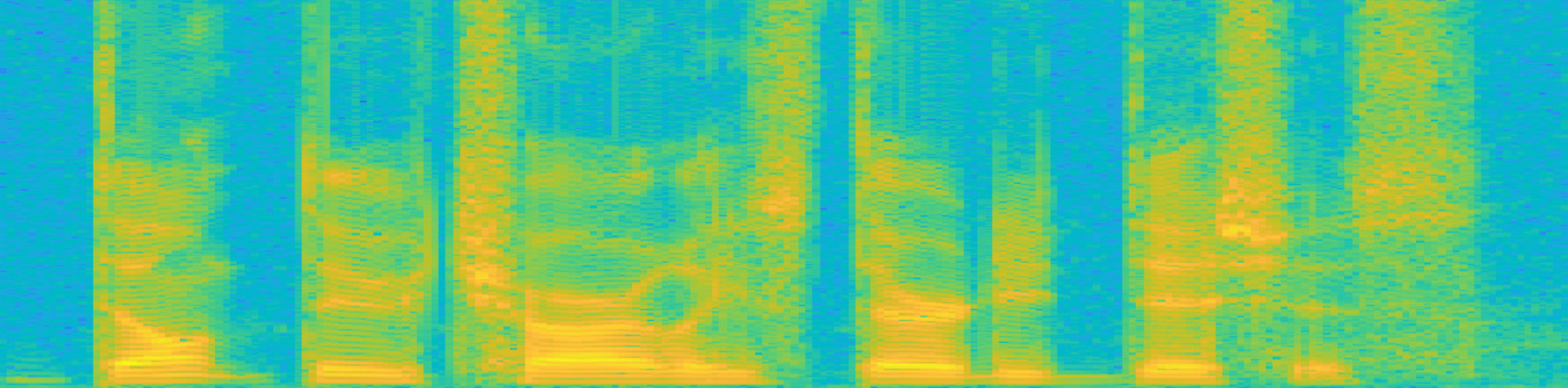} \\
    \includegraphics[height=1.0in,width=0.95\columnwidth]{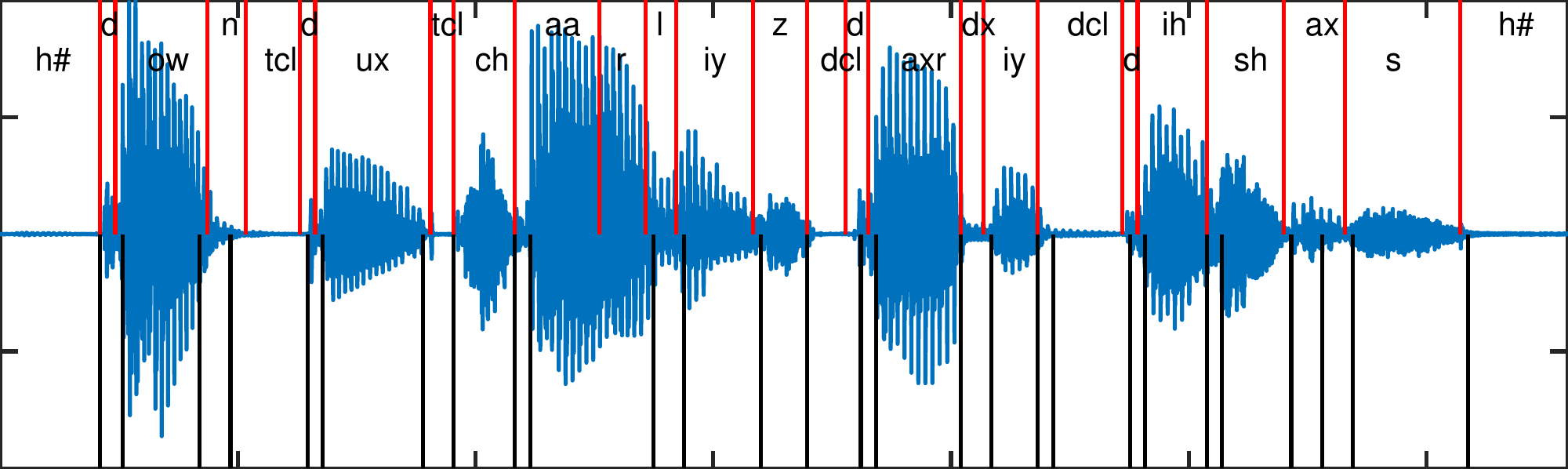} \\
    \includegraphics[height=1.0in,width=0.95\columnwidth]{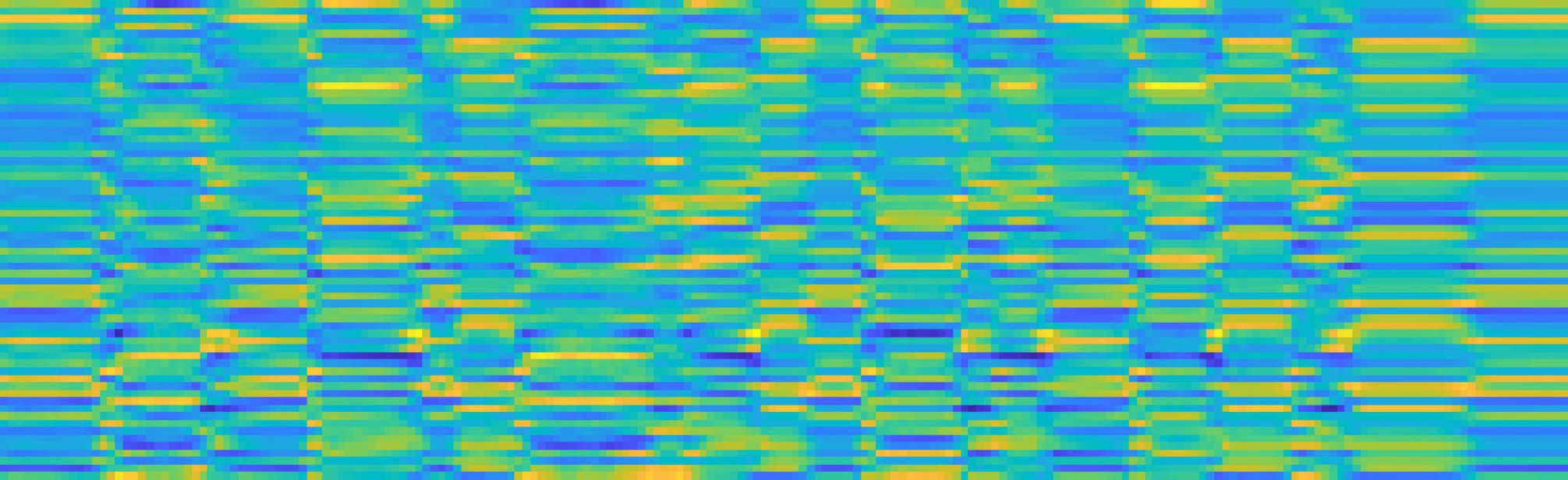}
    \end{minipage} 
    }
    \subfloat[lowest R-val, "he will allow a rare lie" ]{
    \begin{minipage}{0.95\columnwidth}
    \includegraphics[ height=1.0in,width=0.95\columnwidth]{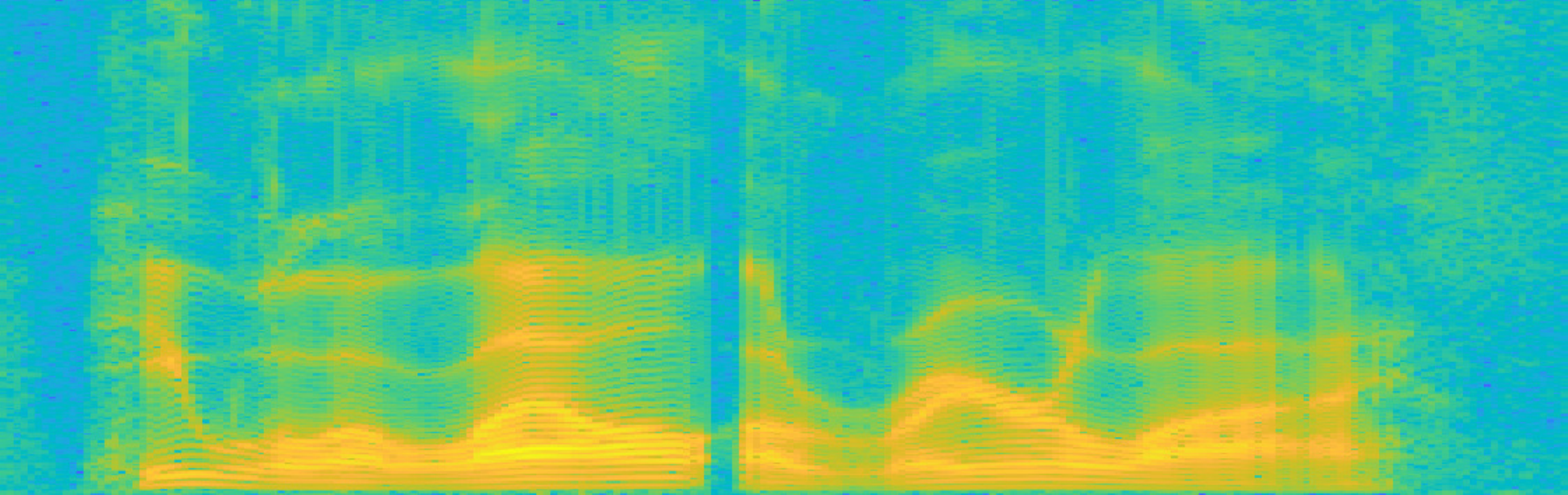}\\
    \includegraphics[height=1.0in,width=0.95\columnwidth]{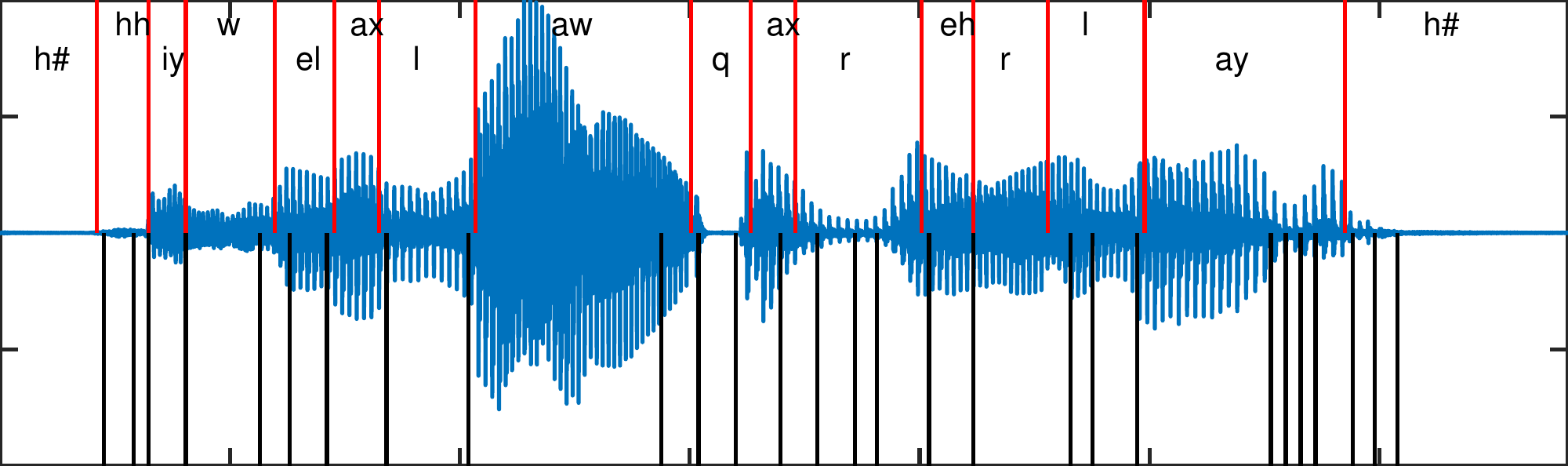}\\
    \includegraphics[height=1.0in,width=0.95\columnwidth]{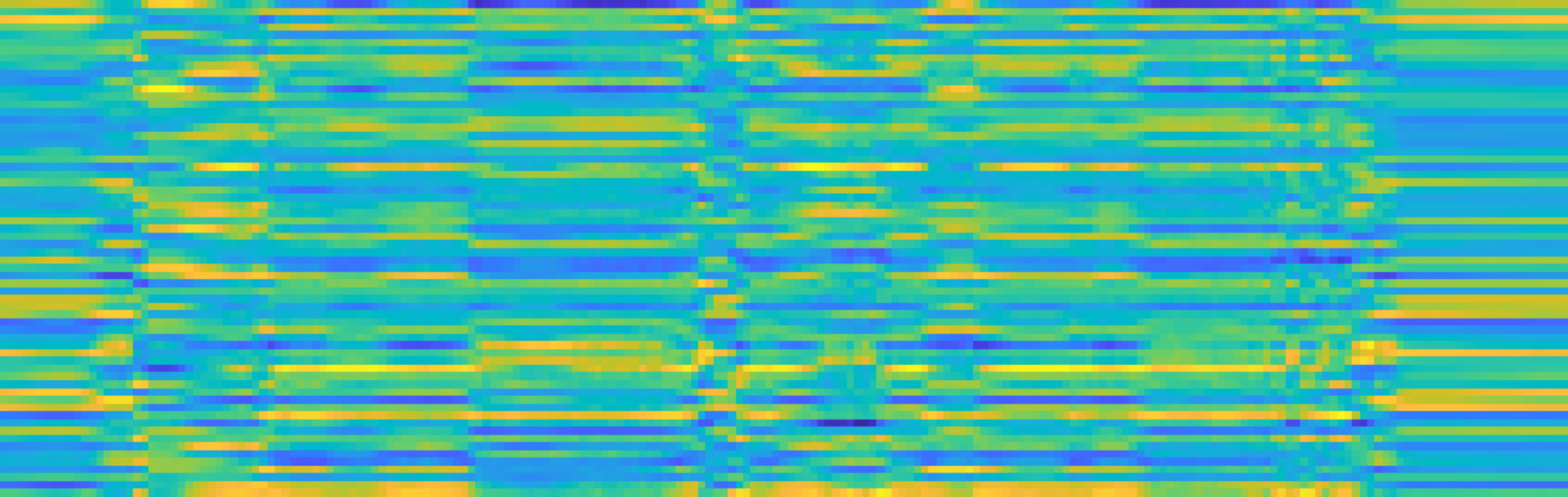} 
    \end{minipage}
    }
    \caption{Spectrogram (top), utterance and the SCPC features, $\mathbf{Z}$, (bottom) of the utterance (best and worst R-val) on the TIMIT dataset. The red line denotes the manual boundaries, and the black lines denote the detected boundaries. The sentences are "don't do charlie's dirty dishes" and "he will allow a rare lie" respectively.    
    }
    \label{fig:best_worst_wav}
\end{figure*}

\subsection{Next Segment Classifier}
The segment encoder $\mathrm{s_{\mathrm{enc}}}$ takes the segment averages as input and outputs the segment representations. 
We use a recurrent neural network (RNN), $s_\mathrm{ar}: \mathbf{S} \rightarrow \mathbf{C} $, to build a contextual representation $(\cvi{1},\cvi{2},...,\cvi{M})$ computed as $c_i = s_\mathrm{ar}(s_{i})$.
Given a reference representation $\cvi{t}$ the model needs to identify the next segment  $\svi{t+1}$ correctly from a set of $K+1$ candidate representations, $\Tilde{\svi{}} \in \mathcal{S}_{t}$ which includes $\svi{t+1}$ and $K$ distractors. 
\begin{equation}
    \mathcal{L}_{\mathrm{NSC}} = -\log \frac{\exp(\mathrm{sim}(\cvi{t},\svi{t+1}))}{\sum_{\Tilde{\svi{}} \in \mathcal{S}_{t} } \exp(\mathrm{sim}(\cvi{t},\Tilde{\svi{}} ))}
\end{equation}
where $\mathrm{sim}$ denotes the cosine similarity.
\subsection{Inference}
During inference, for a new utterance $\mathbf{X}$ we first extract the frame-level features, $\mathbf{Z}$, using $\mathrm{f_{\mathrm{enc}}}$.
For phone segmentation, the model outputs the dissimilarity between adjacent frames. The frames with high dissimilarity are considered segment boundary candidates.  Similar to~\cite{wang2017gate,michel2016blind,kreuk2020self}, we apply a peak detection algorithm to find the final segment boundaries. The peak prominence value for the peak detection algorithm is fined-tuned on the validation dataset~\cite{kreuk2020self}. For word segmentation, the model outputs a dissimilarity score between context representation and segment representation. This can be interpreted as how well the model predicts $\svi{t+1}$ given past context. Segments within a word have higher prediction probabilities than segments across a word boundary~\cite{harris1954distributional,saffran1996word}. A peak detection algorithm locates the times with lower prediction scores which are used as word boundary candidates. Figure \ref{fig:best_worst_wav} shows the input audio utterance, spectrograms, the features learned by the SCPC model, and the manual and predicted phone boundaries. As seen from the figures, the features within a segment are very consistent, and the SCPC features show clear segmentation within them.

\section{Experiments}

\begin{table*}[ht!]
\caption{Comparison of phone segmentation performance on TIMIT and Buckeye test sets. All the results use 20 ms tolerance window.}
\label{tab:phone}
\vspace{-2mm}
\centering
\begin{tabular}{@{}lllll|llll@{}}
\toprule
& \multicolumn{4}{c|}{TIMIT} & \multicolumn{4}{c}{Buckeye} \\
 \cmidrule(l){2-9}
 & Precision & Recall & F1 & \multicolumn{1}{l|}{R-val} & Precision & Recall & F1 & R-val \\ \midrule
RNN ~\cite{michel2016blind} & 74.80 & 81.90 & 78.20 & 80.10 & 69.34 & 65.14 & 67.18 & 72.13 \\
RNN Gate ~\cite{wang2017gate} &  &  &  & \multicolumn{1}{l|}{83.16} & 69.61 & 72.55 & 71.03 & 74.83 \\
CPC ~\cite{kreuk2020self} & 83.89 & 83.55 & 83.71 & \multicolumn{1}{l|}{86.02} & 75.78 & 76.86 & 76.31 & 79.69 \\
CPC+ ~\cite{kreuk2020self} & 84.11 & 84.17 & 84.13 & 86.40 & 74.92 & \textbf{79.41} & 77.09 & 79.82 \\
SCPC & \textbf{84.63} & \textbf{86.04} & \textbf{85.33} & \multicolumn{1}{l|}{\textbf{87.44}} & \textbf{76.53} & {78.72} & \textbf{77.61} & \textbf{80.72} \\ \bottomrule
\end{tabular}
\vspace{-4mm}
\end{table*}

\subsection{Datasets and Evaluation Metrics}

We evaluated the proposed model on both TIMIT~\cite{garofolo1993timit} and Buckeye~\cite{pitt2005buckeye} datasets. For the TIMIT dataset, the standard train/test split was used, with $10\%$ of the train data randomly sampled employed as the validation subset. 
We split the data into 80/10/10 for train, test, and validation sets for Buckeye, similarly to~\cite{kreuk2020self}. The longer recordings were divided into smaller ones by cutting during untranscribed fragments, noises, and silences to ease training. Each smaller sequence started and ended with a maximum of 20 ms of non-speech.
To make results more comparable to~\cite{kamper2020towards}, we also trained our word segmentation system on the English training set from the ZeroSpeech 2019 Challenge~\cite{dunbar2019zero} and test on Buckeye. ZeroSpeech English set contains around 15 hours of speech from over 100 speakers. 
Using two different training and evaluation datasets allows us to analyze how well our method generalizes across corpora. 

For both the phone and word segmentation tasks, we measure the performance with precision (P), recall (R), and F-score with a tolerance of 20 ms~\cite{michel2016blind,wang2017gate,kreuk2020self}. F-score is not sensitive enough to capture the trade-off between recall and Over Segmentation (OS), defined as $R/P - 1$, i.e., even a segmentation model that predicts a boundary every 40 ms can achieve a high F1-score by maximizing recall at the cost of low precision. This motivated a more robust metric, R-value~\cite{rasanen2009improved}, more sensitive to OS, and the only way to obtain a perfect score (1) is to have perfect recall (1) and perfect OS (0). 

\begin{equation}
    \textrm{R-value} = 1 - \frac{\vert r_1\vert + \vert r_2\vert }{2}
\end{equation}
\begin{equation}
    r_1 = \sqrt{(1 - R)^2 + \textrm{OS}^2}, r_2 = \frac{-\textrm{OS} + R - 1}{\sqrt{2}}
\end{equation}

All the results reported here are an average of three different runs with different seeds unless stated otherwise.

\subsection{Model Architecture}\label{Model_arch}
The function $f_\mathrm{enc}$ was implemented using a deep convolutional neural network. It contains five layers of 1-D strided convolution with kernel sizes of (10,8,4,4,4) and strides of (5,4,2,2,2) and 256 channels per layer. The total downsampling rate of the network is 160. 
Each convolution layer is followed by Batch-Normalization and Leaky ReLU non-linear activation function. A fully connected layer linearly projects the outputs in a smaller dimension, 64. Overall, the $f_\mathrm{enc}$ architecture is same as~\cite{kreuk2020self}. Similar to~\cite{kreuk2020self}, SCPC does not utilize a context network at the frame level.  

The segment encoder $s_{enc}$ is a two-layer feed-forward network with 256 neurons each. The auto-regressive network consists of a GRU with 64 neurons followed by a linear layer with 256 neurons. Overall the model has around 1.4 million parameters, and the majority of the model parameters are in the next frame classifier. The model is optimized using a batch size of 8 utterances. We use Adam optimizer with a learning rate of 1e-4 for 100 epochs. By default, the threshold for the boundary detector is set at $0.05$, and the NSC loss is added after two epochs.

\subsection{Effect of Segmental CPC on phone segmentation}

As observed from Table \ref{tab:phone}, SCPC with average as the segment representation outperformed all the baselines. Our proposed approach, SCPC, extracts more structure from the speech data, i.e., at frame-level and at phone-level, whereas the baselines rely on just the frame-level structure. 

The CPC+ line denotes the CPC system trained with additional unlabelled training data from Librispeech~\cite{kreuk2020self}. For TIMIT, the $100$ hours partition and for Buckeye $500$ hours partition are added to their respective training sets. SCPC outperforms the CPC trained with just the TIMIT and Buckeye datasets and 
with data augmentations from Librispeech.

\subsection{Word segmentation performance}
The word segmentation performance on the Buckeye dataset is shown in Table~\ref{tab:word}. 
SCPC clearly outperforms
both the neural~\cite{kamper2020towards} and non-neural~\cite{kamper2017embedded,kamper2017segmental} methods for word segmentation. 
Unlike ES K-Means~\cite{kamper2017embedded} and BES GMM ~\cite{kamper2017segmental}, our proposed method does not require an initial segmentation method and can generate and adjust the boundaries during the learning process.
VQ-CPC and VQ-VAE are pretrained without the segmentation task~\cite{van2020vector} and then optimized for word segmentation, while SCPC is trained jointly. Our model is implicitly encouraged to assign blocks of feature frames to the same segment during the learning process. These approaches have different steps for feature extraction, feature learning, initial segmentation, and then word segmentation. In contrast, in our case, everything is done by a single model jointly with feedback from each other to improve performance. Although both VQ-CPC and VQ-VAE based models have higher recall than our system, it is achieved by over-segmenting the data, which is indicated by the very high OS in the table.

Buckeye\_SCPC denotes the SCPC system trained on Buckeye dataset and tested on same, and 
ZS\_SCPC denotes SCPC system trained on Zerospeech 2019 dataset. 
The validation and test performance are very close, which shows that the model is generalizing well. 
The high performance across datasets shows the robustness of this approach. The slight difference in performance is because ZS\_SCPC is trained on more data. 

\begin{table}[]
\caption{Word boundary segmentation performance on Buckeye development dataset. Parenthesis shows performance on the test set. All the results use a 20 ms tolerance window. 
}
\vspace{-2mm}
\label{tab:word}
\resizebox{\columnwidth}{!}{
\begin{tabular}{@{}llllll@{}}
\toprule
Model & P & R & F1 & OS & R-value \\ 
\midrule
ES K-Means~\cite{kamper2017embedded} & 30.7 & 18.0 & 22.7 & -41.2 & 39.7 \\
BES GMM~\cite{kamper2017segmental} & 31.7 & 13.8 & 19.2 & -56.6 & 37.9 \\
VQ-CPC DP~\cite{kamper2020towards} & 15.5 & \textbf{81.0} & 26.1 & 421.4 & -266.6 \\
VQ-VAE DP~\cite{kamper2020towards} & 15.8 & 68.1 & 25.7 & 330.9 & -194.5 \\
AG VQ-CPC DP~\cite{kamper2020towards} & 18.2 & 54.1 & 27.3 & 196.4 & -86.5 \\
AG VQ-VAE DP~\cite{kamper2020towards} & 16.4 & 56.8 & 25.5 & 245.2 & -126.5 \\
ZS\_SCPC & \textbf{37.3} & 32.7 & \textbf{34.9} & -12.3 & \textbf{46.3} \\
& (36.7) & (31.2) & (33.7) & (-15.2) & (45.8) \\
Buckeye\_SCPC & 35.0 & 29.6 & 32.1 & \textbf{-15.4} & 44.5 \\ 
& (33.3) & (29.7) & (31.4) & (-10.8) & (43.4) \\ 
\bottomrule
\end{tabular}
}
\vspace{-4mm}
\end{table}

\begin{table*}[ht!]
\caption{Fixed vs learning the threshold. Comparison of phone and word segmentation performance on TIMIT and Buckeye test sets. All the results use 20 ms tolerance window.}
\label{tab:FixVsLearnThres}
\vspace{-2mm}
\centering
\begin{tabular}{@{}llllll|llll@{}}
\toprule
& & \multicolumn{4}{c|}{TIMIT} & \multicolumn{4}{c}{Buckeye} \\
 \cmidrule(l){2-10}
&  & Precision & Recall & F1 & \multicolumn{1}{l|}{R-val} & Precision & Recall & F1 & R-val \\ \midrule
\multirow{2}{*}{phone} & fixed threshold & \textbf{84.63} & \textbf{86.04} & \textbf{85.33} & \multicolumn{1}{l|}{\textbf{87.44}} & \textbf{76.53} & \textbf{78.72} & \textbf{77.61} & \textbf{80.72} \\
& learned threshold & 84.06 & 85.56 & 84.80 & \multicolumn{1}{l|}{86.94} & 76.33 & 78.12 & 77.22 & 80.43 \\ \midrule

\multirow{2}{*}{Word} & fixed threshold & \textbf{30.45} & 19.96 & 24.11 & \multicolumn{1}{l|}{40.31} & 36.70 & 29.31 & 32.59 & 45.36 \\
& learned threshold & 30.33 & \textbf{20.59} & \textbf{24.52} & \multicolumn{1}{l|}{\textbf{40.47}} & \textbf{37.31} & \textbf{32.70} & \textbf{34.85} & \textbf{46.33} \\ \bottomrule
\end{tabular}
\vspace{-4mm}
\end{table*}

\subsection{Effect of boundary threshold}
\begin{figure}[h!]
    \centering
    \subfloat[phone segmentation]{\includegraphics[width=1.7in]{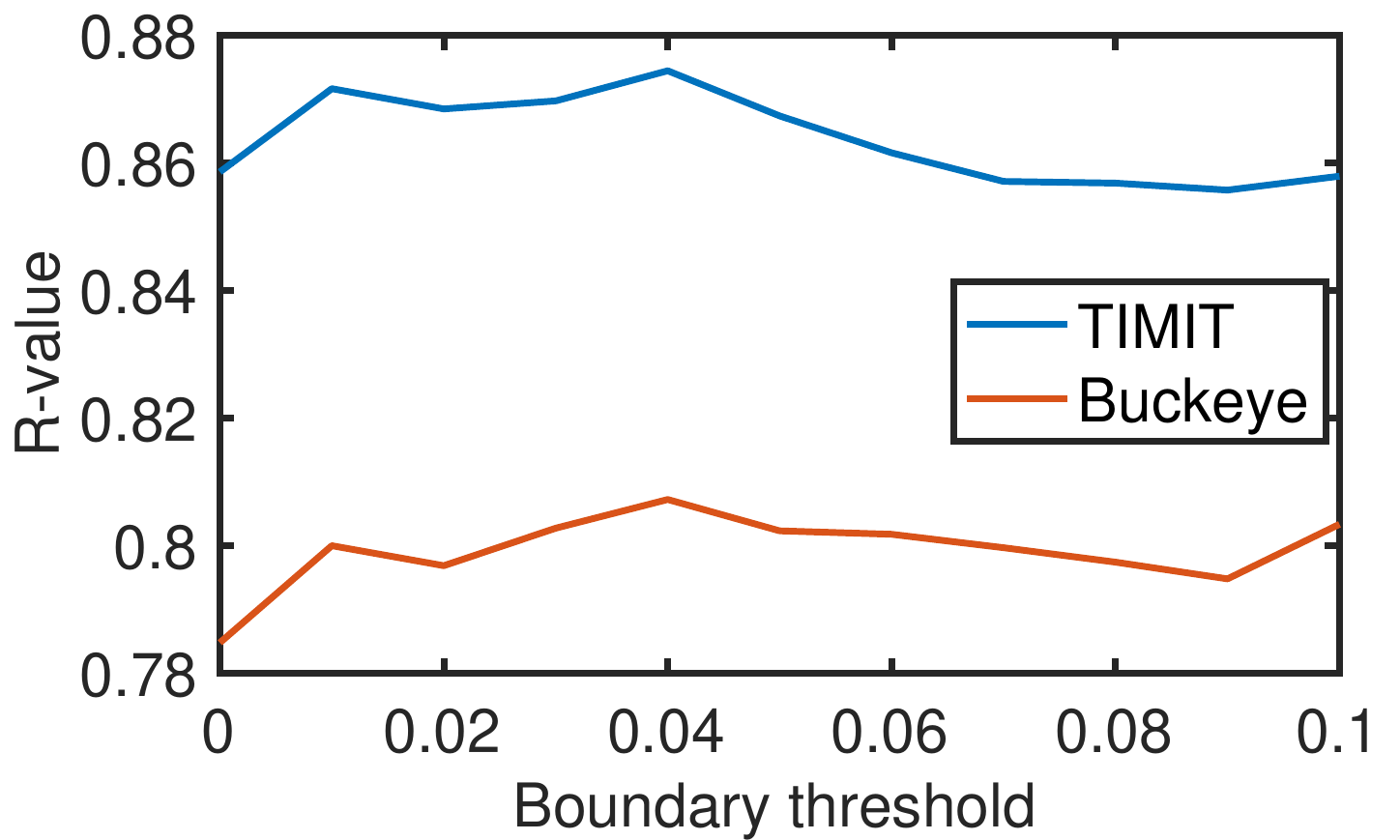}}
    \subfloat[word segmentation]{\includegraphics[width=1.7in]{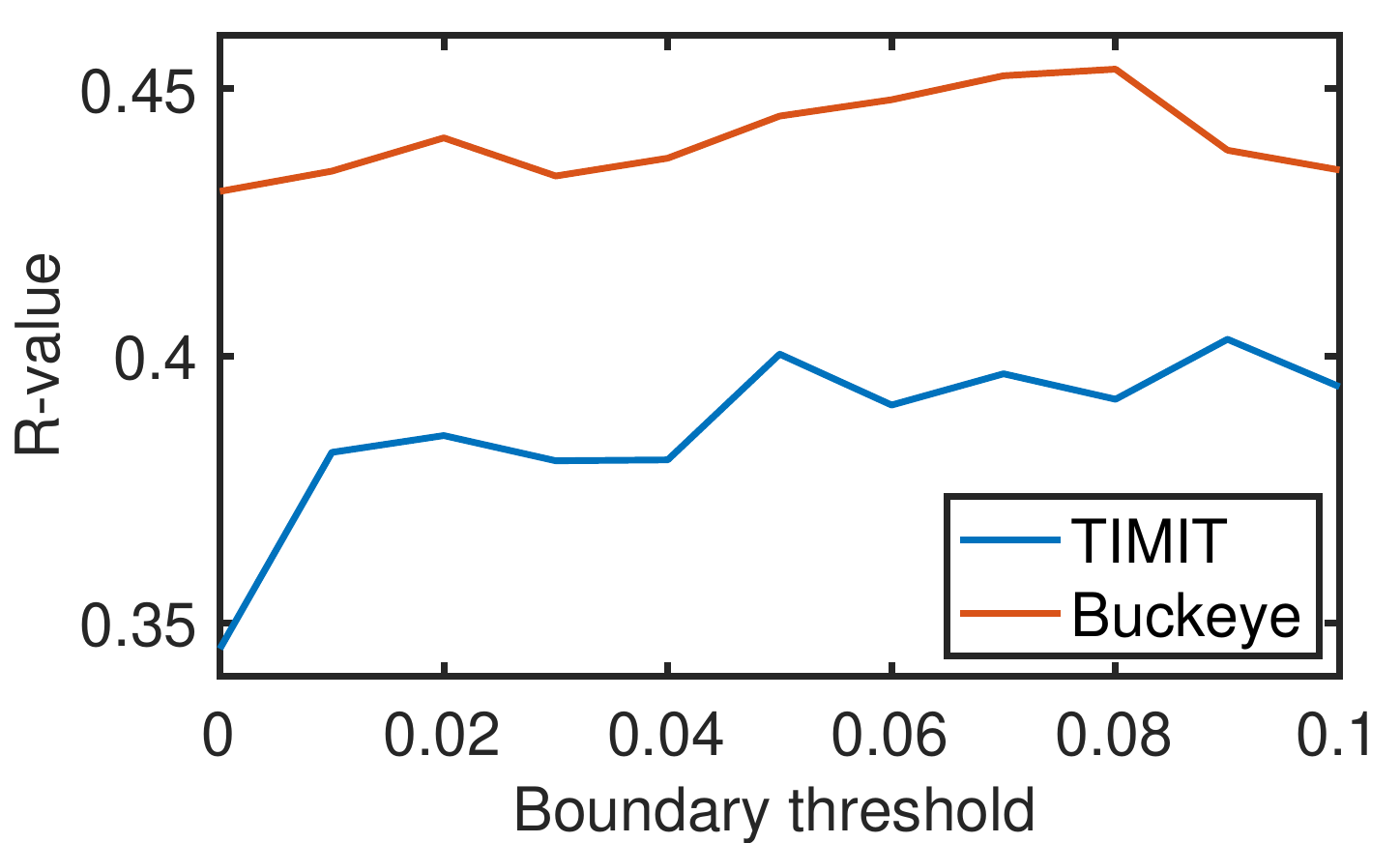}}
    \caption{Segmentation performance on test portions from TIMIT and Buckeye vs fixed boundary threshold value}
    \vspace{-0.2in}
    \label{fig:R_valVsPara}
\end{figure}

The boundary detector uses a threshold to remove the short peaks. The threshold parameter essentially controls the boundary locations, the number of boundaries which in turn dictate the model performance.
To analyze the impact of boundary threshold on system performance, we vary the threshold from $0$ to $0.1$ with 0.01 step size and trained SCPC models. The segmental loss is added after two epochs. 
As observed in Figure~\ref{fig:R_valVsPara} the phone segmentation performs better with lower thresholds. Both datasets achieve the best phone segmentation performance with a peak threshold of $0.04$. The optimal threshold for the word segmentation is higher around $0.09$ and $0.08$ for TIMIT and Buckeye respectively. 
We hypothesize that a lower peak threshold allows the model to generate more boundaries which can match better with the higher number of ground truth phone boundaries present where word boundaries are fewer, so a higher peak threshold suppresses the unnecessary boundaries for improved performance. 

\subsection{Learnable threshold}

Figure~\ref{fig:R_valVsPara} shows the impact of threshold on the final segmentation R-val. 
Our formulation of the boundary detector allows the threshold to be learned with the model weights.
So instead of using a fixed threshold across the entire training process, the model automatically learns the boundary threshold.
Table~\ref{tab:FixVsLearnThres} shows the phone and word segmentation performance for the fixed threshold vs. the learnable threshold. As evident from the table, the performance for the two cases, i.e., fixed vs. learned threshold, is close. 

\begin{figure}[h!]
    \centering
    \subfloat[phone segmentation]{\includegraphics[width=1.7in]{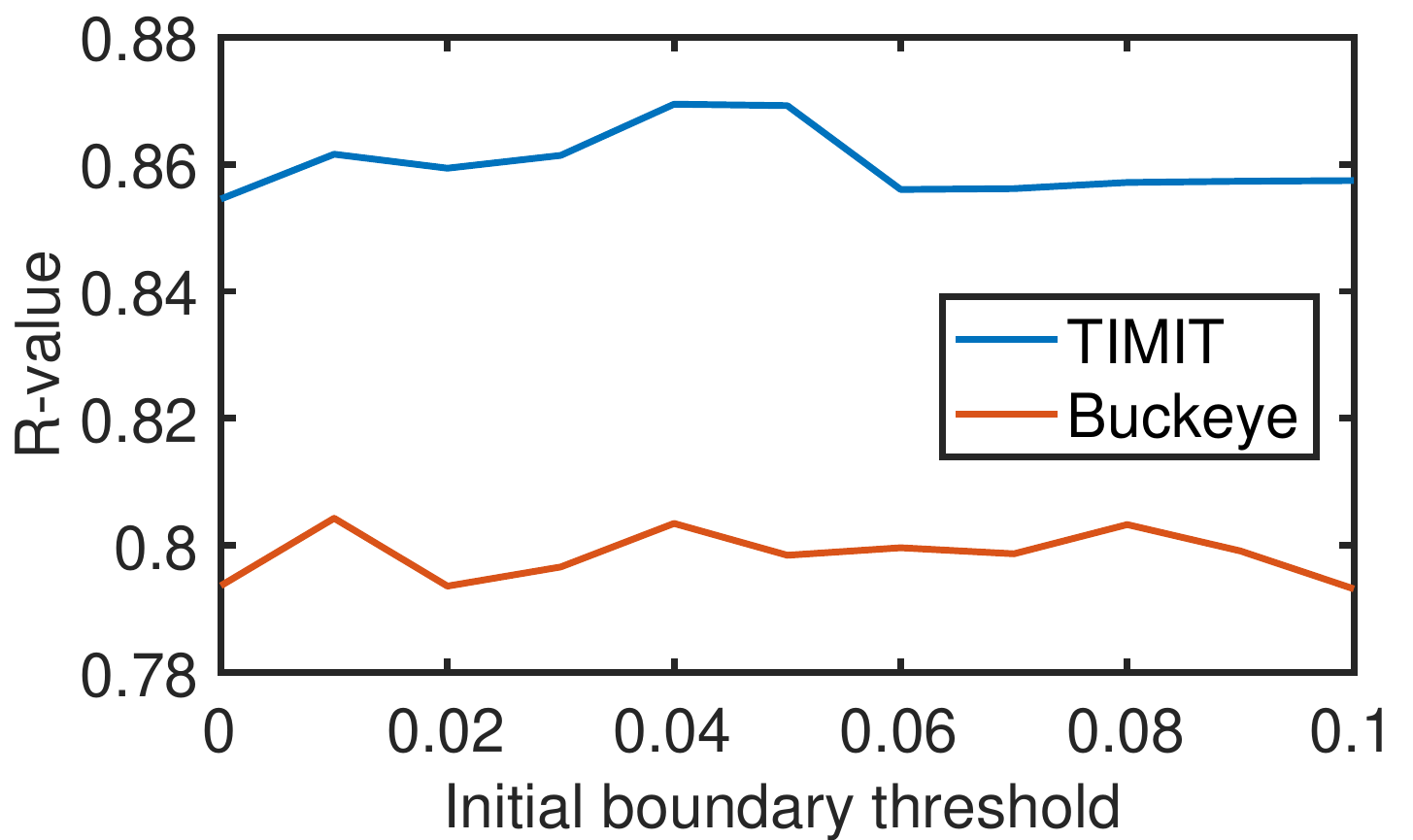}}
    \subfloat[word segmentation]{\includegraphics[width=1.7in]{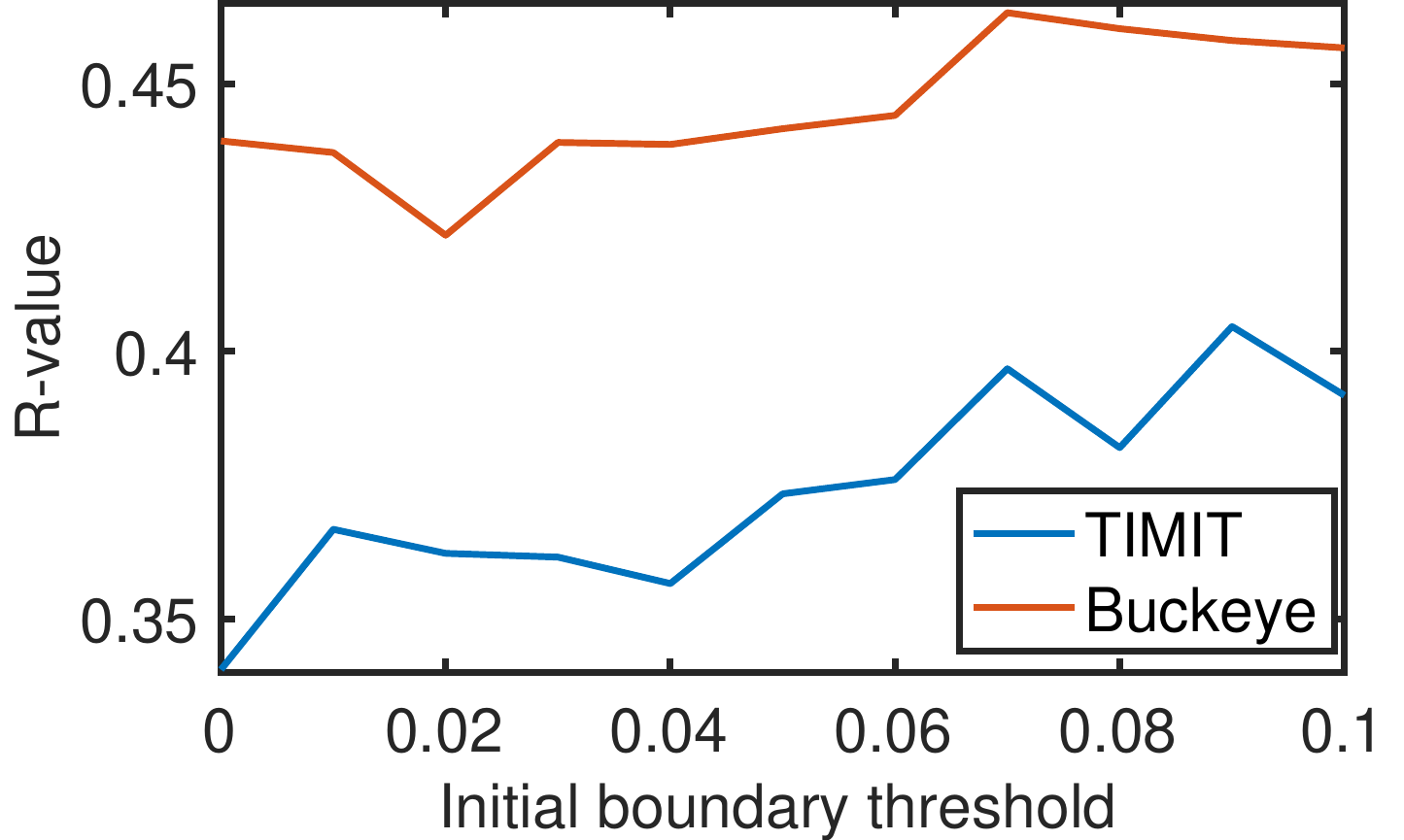}}
    \caption{Segmentation performance on test portions from TIMIT and Buckeye vs the initial value for the learnable boundary threshold}
    \vspace{-0.2in}
    \label{fig:R_valVslthres}
\end{figure}

To analyze the impact of boundary initialization on the system performance, we vary the threshold initialization from $0$ to $0.1$ with 0.01 step size and trained SCPC models. The segmental loss is added after two epochs. Figure~\ref{fig:R_valVslthres} shows the final system performance with varying threshold initializations. The phone performance seems to benefit from lower threshold values and the word segmentation from higher values similar to the Fixed threshold case.

\subsection{When to add NSC loss}
\begin{figure}[h!]
    \centering
    \subfloat[phone segmentation]{\includegraphics[width=1.7in]{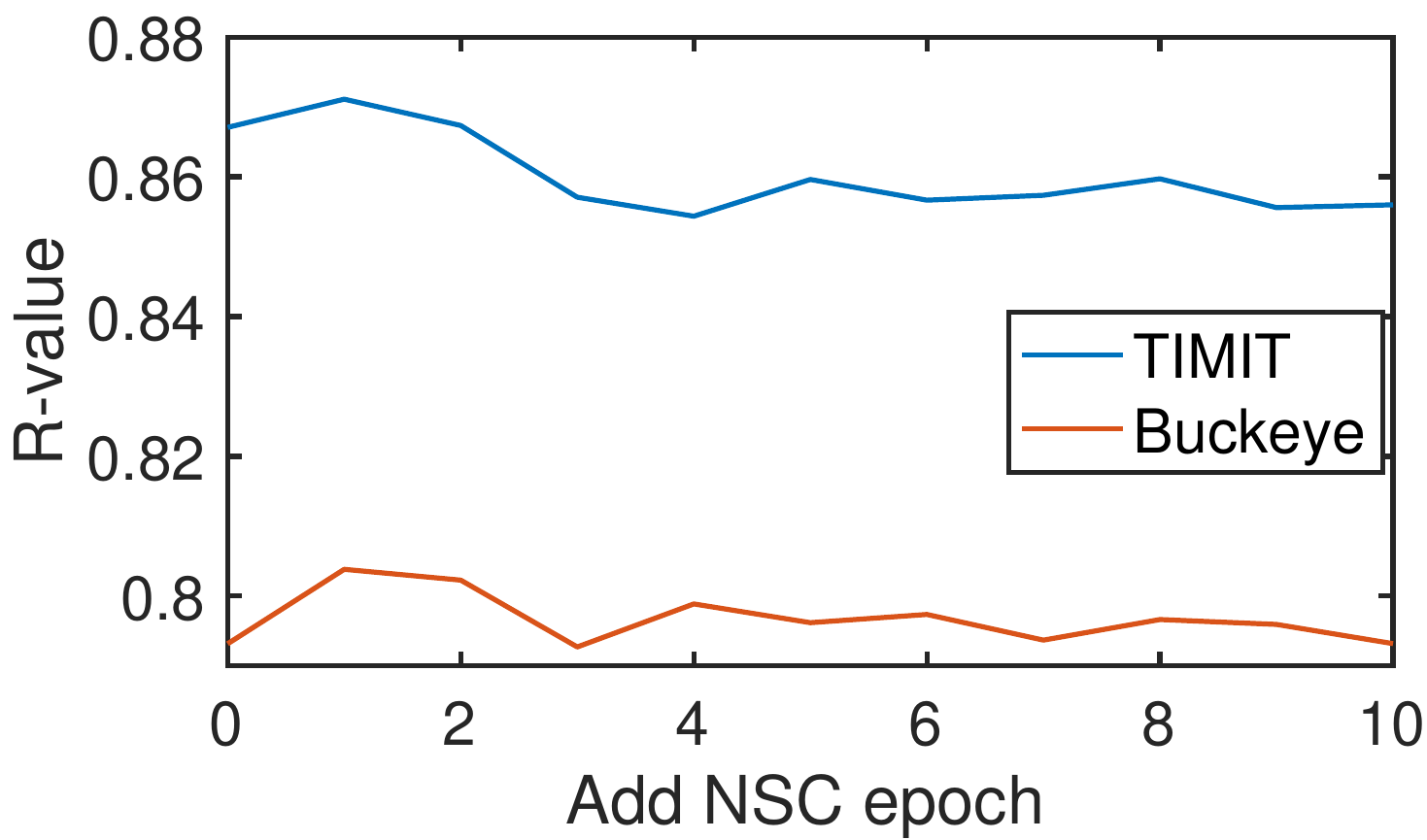}}
    \subfloat[word segmentation]{\includegraphics[width=1.7in]{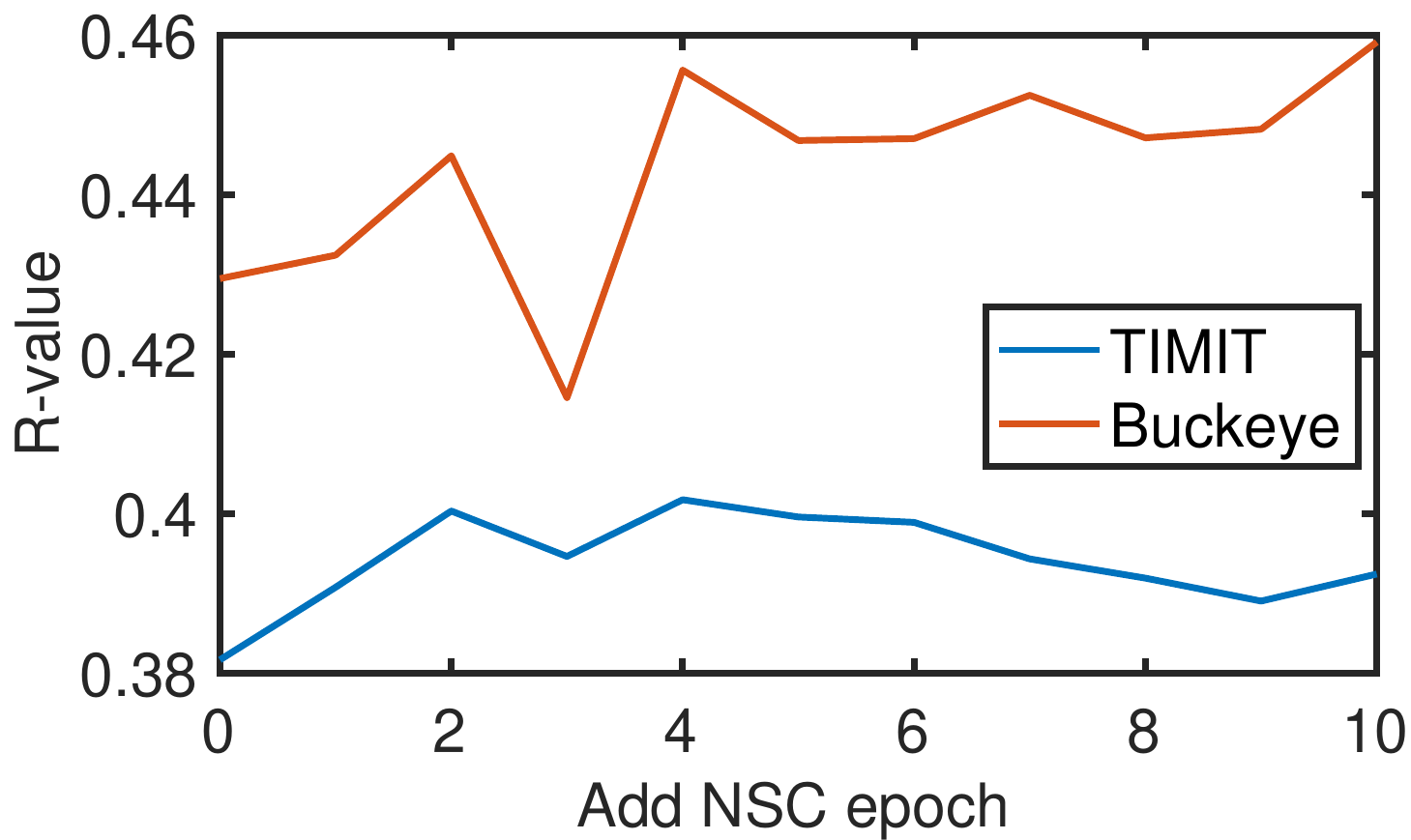}}
    \caption{Segmentation performance on test portions from TIMIT and Buckeye vs epoch at which NSC loss is added}
    \vspace{-0.2in}
    \label{fig:R_valVsadd_nsc}
\end{figure}

To analyze the importance of the epoch at which the segmental loss is added to the model objective, we trained SCPC where the segmental loss is added after $i^{th}$ epoch and $i$ is varied from $0$ to $10$. The boundary threshold is kept at $0.05$. As observed, in Figure~\ref{fig:R_valVsadd_nsc} the phone segmentation performs better if segmental loss is added early ($i = 1$) and word segmentation performs better when segmental loss is added late ($i = 4$). We hypothesize that this allows the model to learn better frame-level features before combining them into segments.

\subsection{Impact of using $ p^{(1)}_{t} $ vs $ p^{(1)}_{t} $ and $ p^{(2)}_{t} $}

The boundary detector relies on both $p^{(1)}$ and $p^{(2)}$ to detect the boundary locations. We hypothesized that $p^{(1)}$ in itself is noisy, and combining it with $p^{(2)}$ improves the boundary detection. We conduct the following experiment; we train a SCPC system with just $p^{(1)}$ and another with both $p^{(1)}$ and $p^{(2)}$. As seen from the table~\ref{tab:p1p2} using both outperforms just the $p^{(1)}$. For this experiment, the peak threshold is set to 0.05, and the NFC is added after two epochs. 

\begin{table}[h!]
    \centering
    \caption{Effects of using just $ p^{(1)}_{t} $ vs $ p^{(1)}_{t} $ and $ p^{(2)}_{t} $. R-values for phone and word segmentation task}
    \begin{tabular}{lllll}
    \toprule
     & \multicolumn{2}{c|}{TIMIT} & \multicolumn{2}{c}{Buckeye}\\ \cmidrule(l){2-5}
    & phn & \multicolumn{1}{l|}{wrd} & phn & wrd \\ \midrule
    $ p^{(1)}_{t} $ & 85.69 & \multicolumn{1}{l|}{36.18} & 78.15 & 40.8\\
    $ p^{(1)}_{t} $ and $ p^{(2)}_{t} $ & \textbf{86.73} & \multicolumn{1}{l|}{\textbf{40.03}} & \textbf{80.23} & \textbf{44.49} \\ \bottomrule
    \end{tabular}
    \label{tab:p1p2}
\end{table}

\begin{table*}[ht!]
\caption{Effect of different types of segment representation on phone and word segmentation on TIMIT dataset.}
\centering
\label{tab:SegRep}
\begin{tabular}{@{}lllll|llll@{}}
\toprule
& \multicolumn{4}{c|}{phone Segmentation} & \multicolumn{4}{c}{Word Segmentation} \\
 \cmidrule(l){2-9}
Seg Representation & Precision & Recall & F1 & R-value & Precision & Recall & F1 & R-value \\ \midrule 
Avg & 84.3 & 84.74 & 84.51 & 86.73 & 28.95 & 23.02 & 25.58 & 40.03 \\
Max & 83.32 & 84.29 & 83.8 & 86.13 & 25.48 & 18.7 & 21.55 & 37.85 \\
Mid & 82.95 & 83.56 & 83.24 & 85.6 & 22.58 & 14.32 & 17.52 & 36.05 \\
WAvg & 83.41 & 84.21 & 83.81 & 86.14 & 27.94 & 18.99 & 22.61 & 39.13 \\ \bottomrule
\end{tabular}
\end{table*}

\subsection{Negative sampling}
We compare different strategies for drawing distractors or negative examples. Table~\ref{tab:NegSamp} shows the R-values on phone and word segmentation tasks. We observe that drawing negative examples from the same utterance performs better than drawing from mixed utterances. This observation is similar to \cite{oord2018representation}, sampling from the same utterance performs better at the downstream phone classification task than the mixed utterance sampling. 

\begin{table}[]
    \centering
    \caption{Effects of using sampling negative examples from same or different utterances}
    \begin{tabular}{lllll} 
    \toprule
     & \multicolumn{2}{c|}{TIMIT} & \multicolumn{2}{c}{Buckeye}\\ \cmidrule(l){2-5}
    & phn & \multicolumn{1}{l|}{wrd} & phn & wrd \\ \midrule
    Mixed-utterance & 86.3 & \multicolumn{1}{l|}{37.44} & 71.96 & 41.21\\
    Same-utterance & \textbf{86.73} & \multicolumn{1}{l|}{\textbf{40.03}} & \textbf{80.23} & \textbf{44.49} \\ \bottomrule
    \end{tabular}
    \label{tab:NegSamp}
\end{table}

\subsection{Context aggregator network }

The next segment prediction task uses a recurrent network to aggregate context information at the segment level. We want to analyze if the previous segment is enough to predict the next segment or more information is required. Like the next frame prediction task, we experiment with predicting the next segment from just the previous segment instead of a recurrent network. As seen in table~\ref{tab:bigramVSrnn}, we observe a performance drop with just using the previous segments. The performance drop is more noticeable for the word segmentation than the phone segmentation. 

\begin{table}[h!]
    \centering
    \caption{Effects of using Bi-gram or RNN}
    \begin{tabular}{lllll}
    \toprule
     & \multicolumn{2}{c|}{TIMIT} & \multicolumn{2}{c}{Buckeye}\\ \cmidrule{2-5}
    & phn & \multicolumn{1}{l|}{wrd} & phn & wrd \\ \midrule
    Previous segment & 85.55 & \multicolumn{1}{l|}{38.6} & 79.41 & 41.34 \\
    RNN & \textbf{86.73} & \multicolumn{1}{l|}{\textbf{40.03}} & \textbf{80.23} & \textbf{44.49} \\ \bottomrule
    \end{tabular}
    \label{tab:bigramVSrnn}
\end{table}  

\subsection{Effect of Segment Representation}

We experiment with the following segment representations: Average, max, mid, and weighted attention. The average-based representation performs the best, followed closely by the weighted average, as observed in Table~\ref{tab:SegRep}. The average-based representation is also the most straightforward to extract in a batch manner.  
The segment middle-point-based representation performs the worst amongst the four. 

Our hypothesis is noise in the boundary detection, and the evaluation process might be responsible for this. For e.g., consider a phone "iy" and three instances of the same phone 1) boundary is detected exactly at the manual boundary 2) the boundary is detected two frames after the manual boundary 3) the boundary is detected two frames before the manual boundary. 
In all three cases, the segmentation performance is unaffected ( as the threshold for the boundary detection is 20 ms or two frames), but the middle point of the segment will be different in the three cases.
This might lead to a different segment representation.
The maximum of the segment might or might not change depending on the next segment. The average or weighted average is least affected by the inclusion/exclusion of one or two frames as most frames remain the same.

We also experimented with a combination of segment representation, i.e., concatenation of the representation from two or more as the segment representation. The combinations with "avg" as one of the constituents performed better than other combinations. However, the performance was still lower than avg as the segment representation.

\section{Boundary detection analysis}

The boundary detection performance might vary depending on the class labels of the successive phones in between which we are trying to identify the boundary. We propose the following experiment to determine the phone classes between which the boundaries are most challenging to locate. We take a pair of two consecutive phone segments and count how many times a predicted boundary falls within 20 ms of the manual boundary between the phones. We divide this count by the total number of manual boundaries for that pair of phones to obtain a final score.  

We can repeat this experiment on all pairs of successive phones. This results in a 61x61 table. To make the results more comprehensible, we use the manual phones label and convert them into broad phonetic labels. The Table~\ref{tab:BFC_labels_2} shows the mapping between the Broad phonetic classes and phonetic labels~\cite{shon2018frame}. We experiment with the broad phonetic labels of the phones. 

\begin{table}[h!]
    \centering
    \caption{Broad Phonetic classes and constituting phone labels in TIMIT.}
    \label{tab:BFC_labels_2}    
    \begin{tabular}{|c|c|}
         \hline
        Broad-phonetic class & phones \\ \hline
        Affricates & jh,ch \\ \hline
        Closures & bcl,dcl,gcl,pcl,tcl,kcl \\ \hline
        Voiceless Fricatives & s,sh,f,th,hv \\ \hline
        Voiced Fricatives & z,zh,v,dh \\ \hline
        Nasals & m,n,ng,em,en,eng,nx \\ \hline
        Semivowels & l,r,w,y,el \\
        (and other voiced consonants) &  \\ \hline %
        Vowels & iy,ih,eh,ey,ae,aa,aw,ay,ah,ao,oy, \\ 
        & ow,uh,uw,ux,er,ax,ix,axr,ax-h \\ \hline
        Voiceless Stops & p,t,k,dx,q \\ \hline
        Voiced Stops & b,d,g \\ \hline
        Others & pau,epi,h\# \\ \hline
    \end{tabular}
\end{table}

\begin{table*}[]
\caption{Successive phone boundary detection accuracy (\%) matrix for CPC based segmentation on TIMIT test set. All the results use 20 ms tolerance window.}
\label{tab:cpc_analysis}
\begin{tabular}{lllllllllll}
\toprule
 & Affricate & Closures & VoicelessFricative & VoicedFricative & Nasals & Semivowels & Vowels & VoicelssStops & VoicedStops & Others \\
\cmidrule(l){2-11}
Affricate & - & 86.96 & 22.22 & - & 92.31 & 90.00 & 93.09 & - & - & 73.53 \\
Closures & 95.68 & - & 90.91 & 70.00 & 86.64 & 87.03 & 88.78 & 92.18 & 91.96 & 27.67 \\
VoicelessFricative & - & 93.07 & 58.00 & 71.88 & 89.32 & 95.83 & 94.70 & 100.00 & - & 75.36 \\
VoicedFricative & - & 92.55 & 64.90 & 75.00 & 91.92 & 90.91 & 94.52 & 93.75 & - & 79.73 \\
Nasals & 78.57 & 58.39 & 88.74 & 70.73 & 17.95 & 67.06 & 89.90 & 87.25 & 91.84 & 45.76 \\
Semivowels & - & 96.60 & 97.76 & 87.90 & 90.67 & 72.95 & 53.32 & 66.67 & - & 83.09 \\
Vowels & - & 96.91 & 93.76 & 92.25 & 89.76 & 42.43 & 31.88 & 80.75 & - & 81.15 \\
VoicelssStops & - & 100.00 & 86.98 & 69.74 & 94.96 & 94.79 & 91.94 & 95.16 & - & 71.89 \\
VoicedStops & - & 93.33 & 87.10 & 100.00 & - & 94.41 & 97.62 & 100.00 & - & 76.32 \\
Others & 100.00 & 70.00 & 78.83 & 84.31 & 95.98 & 98.51 & 96.30 & 96.48 & 97.66 & - \\ \bottomrule
\end{tabular}
\end{table*}

\begin{table*}[]
\caption{Successive phone boundary detection accuracy (\%) matrix for SCPC based segmentation on TIMIT test set}
\label{tab:scpc_analysis}
\begin{tabular}{lllllllllll}
\toprule
 & Affricate & Closures & VoicelessFricative & VoicedFricative & Nasals & Semivowels & Vowels & VoicelssStops & VoicedStops & Others \\
\cmidrule(l){2-11}
Affricate & - & 97.10 & 48.15 & - & 92.31 & 100.00 & 98.39 & - & - & 83.82 \\
Closures & 95.68 & - & 91.15 & 77.04 & 93.55 & 89.73 & 95.92 & 91.57 & 94.14 & 21.34 \\
VoicelessFricative & - & 92.97 & 64.00 & 81.25 & 96.12 & 98.06 & 97.75 & 97.92 & - & 83.67 \\
VoicedFricative & - & 91.69 & 64.24 & 80.43 & 96.97 & 93.05 & 96.27 & 93.75 & - & 86.15 \\
Nasals & 85.71 & 74.24 & 90.34 & 70.38 & 16.67 & 70.92 & 92.16 & 91.18 & 89.80 & 70.51 \\
Semivowels & - & 99.19 & 97.39 & 94.35 & 96.00 & 79.94 & 64.84 & 77.48 & - & 88.24 \\
Vowels & - & 98.08 & 94.70 & 93.74 & 93.38 & 49.78 & 43.80 & 86.75 & - & 85.51 \\
VoicelssStops & - & 91.67 & 85.65 & 73.68 & 94.24 & 98.02 & 94.48 & 95.16 & - & 71.43 \\
VoicedStops & - & 86.67 & 93.55 & 96.43 & - & 97.81 & 98.81 & 100.00 & - & 76.32 \\
Others & 100.00 & 60.00 & 84.50 & 95.42 & 96.59 & 99.58 & 97.69 & 97.83 & 99.12 & - \\ \bottomrule
\end{tabular}
\end{table*}

We repeat the previous experiment for both CPC and SCPC. As seen from Table~\ref{tab:cpc_analysis} and Table~\ref{tab:scpc_analysis} both CPC and SCPC provide high boundary detection accuracy among groups, with accuracy over $90\%$ and with some exceptions. The results suggest that both techniques struggle with boundaries between vowels followed by vowels or semivowels. Figure\ref{fig:best_worst_wav}(a) illustrates an example of a vowel ([aa]) followed by a voiced consonant ([r]) in the utterance with the highest R-value, where the SCPC misses the boundary between the two phonetic units.

The reason for the low performance in the boundary detection between vowels or vowels and semivowels is that the transition between these sounds is often gradual. It involves changes in the position of articulators such as tongue, lips, and jaw but without clearly defined stops or voiceless turbulence, which makes accurate boundary detection difficult. A similar phenomenon occurs with the transition between two nasals, where the boundary detection is below $18\%$ for CPC and SCPC. It is noticeable that the boundaries between vowels (or semivowels) and nasals are usually well detected, in spite that both types of sounds are voiced, and the transition does not include stops or turbulence. However, in this case, the transition between sounds involves articulating the velopharyngeal port to switch between the oral and nasal cavities, which leads to a better boundary detection due to the acoustic differences between sounds caused by the different cavity resonances. In the same manner, the transition between vowels and voiced fricatives or stops -and vice-versa- is usually well detected with the accuracy normally over 85\%. In these cases, the presence of stops and turbulence created by the narrowing (fricatives) or total closure (stops) of the vocal tract leads to a better differentiation of boundaries, in spite of all the sounds being voiced. We observe that the performance of SCPC is better than for CPC in most of the transitions. We hypothesize that this is because of the context captured by recurrent $s_{ar}$ in the Next Segment Classifier and the model trying to learn information at multiple levels, i.e., frame and segment level.  

\section{Using SCPC for variable-rate representation learning}
From hand-crafted features like MFCC to neural representation like CPC, most feature extraction methods use a fixed sampling rate, i.e., a feature is extracted every 10 ms. The number of features depends upon the duration of the signal rather than on the semantic content, e.g., a one-second silence will be encoded with the same number of features as a 1-second speech signal. This leads to redundancy in the features, and adjacent features capture similar information. An approach to minimize the redundancy would be to sample representation whenever the semantic content in an utterance changes. But this requires us to find locations where the semantic information changes. 

SCPC can divide a given utterance into a variable number of segments. Each segment boundary dictates information change, e.g., the underlying phone has changed. The number of the segment depends on the underlying phone content in the utterance. SCPC then generates a segment representation. Essentially, SCPC can irregularly sample the speech whenever a segment changes and keeps a single representation per segment. An ideal segment-based representation could match the performance of frame-based representation since it captures all the information provided by the frame-level representation.

\subsection{Contrastive learning for segmentation vs representation learning}
Even though SCPC can generate segment representations or frame-level representations, the SCPC is optimized for speech segmentation and not for learning representations that capture the underlying phone classes. Even the CPC model used for segmentation~\cite{kreuk2020self} optimizes a different version of the loss and uses a different architecture than the one used for learning representations~\cite{oord2018representation}. A few of the key differences between the next frame classifier, CPC for segmentation, and CPC for representation learning are:
\begin{itemize}
    \item Absence of a context network: The frame encoder in SCPC, CPC for segmentation, does not use a frame-level recurrent context network.     
    \item Number of steps predicted: The frame encoder in SCPC, CPC for segmentation predict just the next step whereas CPC for representation learning predict much further into the future, 12 steps 
    \item No regression layers: CPC for representation learning uses regression layers to map the output of the context network to the encoder output.
    \item Number of negative examples: The frame encoder in SCPC, CPC for segmentation, uses one negative example whereas CPC for representation learning uses many negative examples, 100. Increasing the number of negative examples decreases the segmentation performance. Authors in~\cite{kreuk2020self} had similar observations.
\end{itemize}

So we modify the next frame classifier in SCPC to include a frame level context network and predict much further into the future. The frame level encoder, $f_{\mathrm{enc}}:\mathbf{X}\rightarrow\mathbf{Z}$, maps the audio waveform, $\mathbf{X}$ to latent spectral representations, $\mathbf{Z}(\in \mathbb{R}^{ p \times L}) = (\zvi{1},\zvi{2},...,\zvi{L})$. Each p-dimensional vector $\zvi{i}$ corresponds to a 30 ms audio frame extracted with 10 ms shift. A recurrent neural network, $f_\mathrm{ar}: \mathbf{S} \rightarrow \mathbf{C}^{f} $, to build a contextual representation $(\cvi{1}^{f},\cvi{2}^{f},...,\cvi{M}^{f})$ computed as $c_i^f = s_\mathrm{ar}(\zvi{i})$.
Given a reference context representation $\cvi{t}^{f}$ the model needs to identify the next frame  $\zvi{t+1}$ correctly from a set of $K+1$ candidate representations, $\Tilde{\zvi{}} \in \mathcal{Z}_{t}$ which includes $\zvi{t+1}$ and $K$ distractors. 

\begin{equation}
    \mathcal{L}_{NFC} = -\frac{1}{M} \sum_{m=1}^{M} \log \frac{\exp(\zvi{t+m}^{T}W_{m}\cvi{t})}{\sum_{\Tilde{\zvi{}} \in \mathcal{Z}_{t} } \exp(\Tilde{\zvi{}^{T}}W_{m}\cvi{t} )}    
\end{equation}

\begin{figure}[h!]
    \centering
    \includegraphics[width=3.3in]{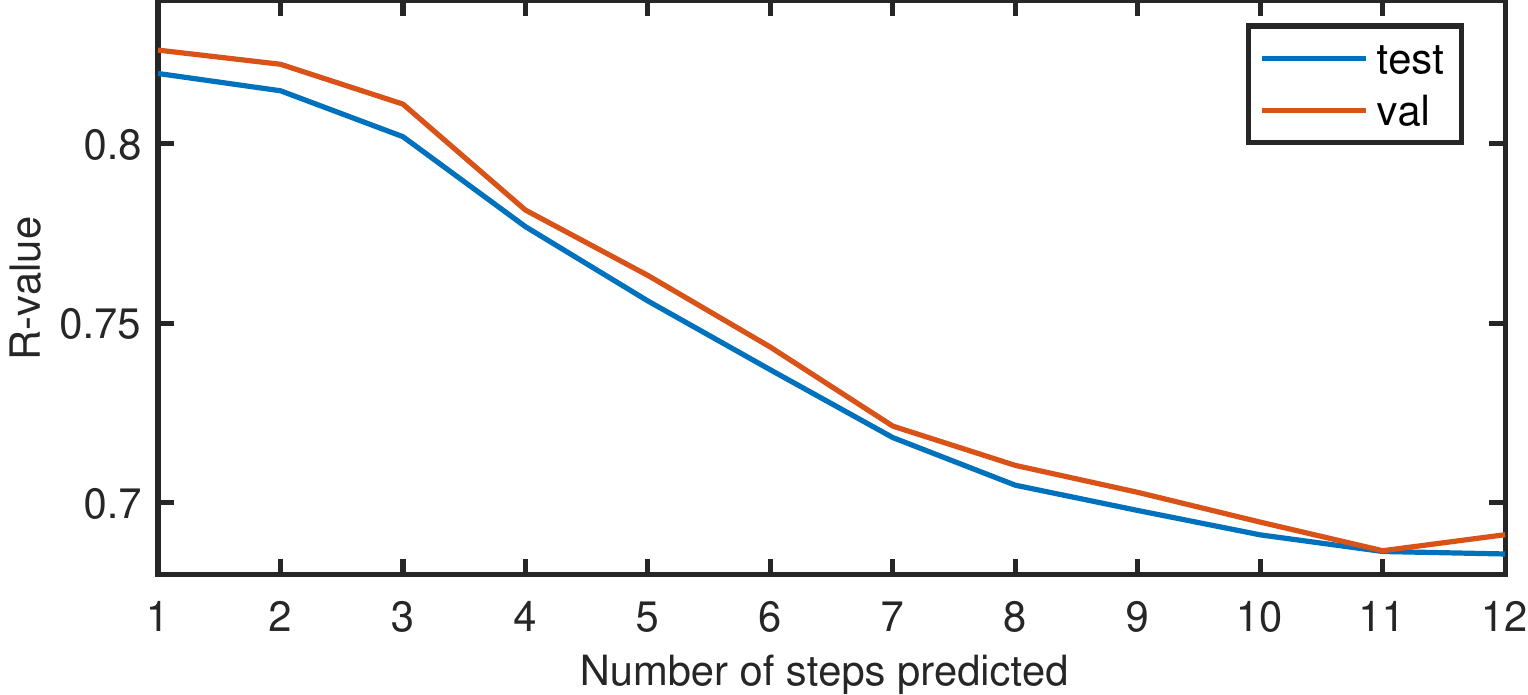}
    \caption{phone segmentation on TIMIT dataset vs the number of steps predicted at frame level. }
    \label{fig:R-val_num-steps}
\end{figure}

As seen in Figure~\ref{fig:R-val_num-steps}, the phone segmentation performance degrades with an increase in the number of steps predicted.  These results seem opposite to the trend in~\cite{oord2018representation} where the linear phone classification performance increases with an increase in the number of steps predicted. As seen in Figure~\ref{fig:R-val_num-steps}, even predicting just the next step with a frame-level context network decreases the segmentation performance. The addition of a frame-level context network reduces the segmentation performance. The observations are similar to CPC for segmentation~\cite{kreuk2020self} whereas it is crucial for learning better representations. 

There seems to be a trade-off between representation learning and
phone segmentation. Unsupervised phone segmentation works better when the model predicts the nearest frame, i.e., the next frame. In contrast, linear phone classification works better when predicting as far into the future as possible. Segmentation relies on the dissimilarity between the adjacent frames, so it makes sense that a model which focuses on predicting the next frame works better than the model which predicts farther into the future.  

\subsection{Two Stage process for extracting Segmental representations}
Instead of using SCPC for both segmentation and representation learning, we divide the two tasks. Figure~\ref{fig:cpc_scpc} shows the overview for the segmental representation learning process. The SCPC still uses a next frame classifier, but instead of taking the raw waveform as input, it takes the features from the pretrained CPC as input. Using the pretrained CPC features as input reduces the boundary segmentation performance from $87$ to $77$ but increases the linear phone classification performance. The architectures of the two components are:

\begin{figure}[h!]
    \centering
    \includegraphics[width=3.2in]{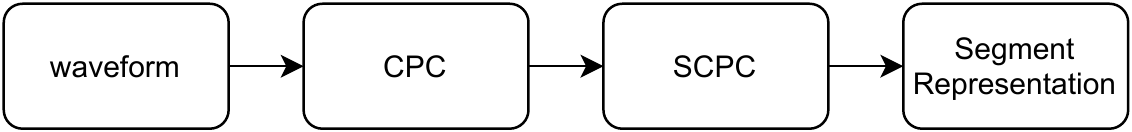}
    \caption{Overview of variable rate segmental representations using SCPC}
    \label{fig:cpc_scpc}
\end{figure}

\begin{itemize}
    \item CPC: We use the embeddings extracted from the pretrained CPC model from~\cite{nguyen2020zero} as input. The model is trained with a context layer, predicts 12 steps into the future. The encoder is a 5-layer 1D-convolutional network with kernel sizes of 10,8,4,4,4 and stride sizes of 5,4,2,2,2 respectively. The model has a downsampling factor of 160, i.e., the embeddings have a sampling rate of 100Hz.
    \item SCPC: The next frame classifier in SCPC is changed from CNN to a feed-forward network with three layers which takes CPC embeddings as input, and the rest of the architecture is the same as before.  
\end{itemize}

We extract the frame-level and segment-level representations from the SCPC. We train a linear classifier to classify phones from the input representations. We repeat each segment's segmental representation length times to align the segmental representations with the per frame phone class levels.  

As seen in Table~\ref{tab:var_rate_rep} results show the segment-level representations outperform the frame-level features, i.e., MFCC on linear phone classification task on TIMIT database, while significantly reducing the sampling rate. There is a decrease in the classification performance from the frame-level features extracted from CPC to segment-level features, which is expected as the information is lost while representing a time-varying segment with a single vector.

The segmentation quality can affect the quality of the segmental representation, which in turn affects the phone classification performance. To analyze this, we segment the frame-level outputs using a non-differentiable peak detector, use the manual boundaries instead of the differentiable peak detector. The gap between the frame-level and segment-level performance decreases as the segmentation improves. Please note that the model is trained with the differentiable boundary detector; manual boundaries are used only during inference.

\begin{table}[]
    \centering
    \caption{Single layer phone classification experiments on TIMIT dataset with 48 phones}    
    \begin{tabular}{lllll}
        \toprule
         & \multicolumn{2}{c|}{Accuracy} & \multicolumn{2}{c}{Average} \\
        	 & \multicolumn{2}{c|}{} & \multicolumn{2}{c}{Sampling Rate} \\ \cmidrule{2-5}
         & val & \multicolumn{1}{l|}{test} & train & \multicolumn{1}{l}{test} \\ \midrule
        MFCC & 47.24 & 47.00 & 100 & 100 \\ \midrule
        frame-level representation ($\mathbf{Z}$) & 67.78 & \multicolumn{1}{l|}{67.19} & 100 & 100 \\ \midrule
        \underline{Segment-level representation ($\mathbf{S}$):} & & & & \\
        Differentiable boundary detector & 58.20 & \multicolumn{1}{l|}{57.59} & 14.47 & 14.27\\
        External Peak detector & 59.58 & \multicolumn{1}{l|}{59.04} & 13.57 & 13.42 \\ 
        Manual boundaries & 62.43 & \multicolumn{1}{l|}{62.45} & 12.07 & 11.99 \\ \bottomrule
    \end{tabular}
    
    \label{tab:var_rate_rep}
\end{table}

\section{Conclusions and future work} 
Segmental Contrastive Predictive Coding (SCPC) paves the way for extending the self-supervised learning framework beyond frame-level auxiliary tasks and unifying the unsupervised phone and word segmentation tasks. SCPC performs unsupervised segmentation of speech into the phone and word-like units by exploiting the speech signal structure at the segment level. SCPC uses a differentiable boundary detector to find variable-length segments (phones). The differentiable boundary detector allows us to pass information between frame and segment levels and jointly optimize frame and segment level representations directly from raw speech waveform. Our experimental results indicate that the learned segments correspond to phone-like units, and the segmental information improves the phone segmentation performance. SCPC outperforms the state-of-the-art phone and word segmentation methods on TIMIT and Buckeye datasets.

We also compared the CPC used for representation learning and CPC used for segmentation. For learning representations, predicting far into the future improves performance, whereas predicting the next frame works best for segmentation. The semantic content in speech varies over time. Fixed-rate representation methods produce features at equally spaced time intervals, and the adjacent features often have redundancies.  We use SCPC for variable rate representation learning and generate representations at the segmental level, i.e., one feature per segment. These representations outperform the fixed-rate MFCC features on the linear phone classification task while significantly lowering the sampling rate.

In the future, we will work to improve the differentiable boundary detector. We would focus on developing efficient algorithms that consider more than two adjacent frames for deciding if a peak exists or not. Experiments show that the threshold for the boundary detector can be learned. Currently, the same threshold is used for all the utterances; utterance-specific thresholds remain to be investigated. We expect that further improvements to the differentiable boundary detector would enhance the performance as it connects the whole system together. We will explore more ways of using unsupervised segmentation for variable-rate representation learning. We also intend to perform downstream task benchmarks of the learned segmental features.

\bibliographystyle{ieeetr}
\bibliography{ref}

\end{document}